\documentclass[twocolumn,secthm]{autart}    

\usepackage{graphicx}          

\usepackage{amsmath,mathrsfs,mathtools,amssymb}
\usepackage{enumerate}

\usepackage{cite}

\newcommand{\mc}[1]{\mathcal{#1}}
\newcommand{\mb}[1]{\mathbb{#1}}
\newcommand{\R}{\mathbb{R}}

\newcommand{\Z}{\mathbb{Z}}
\newcommand{\Sphere}{\mathbb{S}}
\newcommand{\SO}{\mathrm{SO(3)}}

\newcommand{\idQ}{\mathbf{i}}
\newcommand{\SetPoint}{\mathcal{A}}
\newcommand{\SetObject}{\mathcal{B}}
\newcommand{\Warp}{\mathcal{T}}
\newcommand{\indexSet}{\mathbb{Q}}
\newcommand{\CPF}{\mathcal{C}^1_1 (\Sphere^3 \times \indexSet,\R_{\geq 0})}
\DeclareMathOperator{\nullSpace}{\operatorname{null}}
\DeclareMathOperator{\Proj}{\operatorname{Proj}}
\DeclareMathOperator{\domain}{\operatorname{dom}}
\DeclareMathOperator{\grad}{\operatorname{grad}}
\DeclareMathOperator{\argmin}{\operatorname{argmin}}
\DeclareMathOperator{\gradA}{\nabla}
\DeclareMathOperator{\CritPoint}{\operatorname{Crit}}

\DeclarePairedDelimiterX{\Abs}[1]{\lvert}{\rvert}{#1}
\DeclarePairedDelimiterX{\Norm}[1]{\lVert}{\rVert}{#1}

\usepackage{xcolor}

\usepackage{comment}

\includecomment{extendVersion}

\begin{document}

\setlength{\abovedisplayskip}{4pt}
\setlength{\belowdisplayskip}{4pt}

\begin{frontmatter}
\title{Quaternion-Based Attitude Stabilization Using Synergistic Hybrid Feedback With Minimal Potential Functions\thanksref{footnoteinfo}} 

\thanks[footnoteinfo]{Research reported in this work was supported in part by Innovation and Technology Commission of Hong Kong (ITS/136/20, ITS/135/20, ITS/234/21, ITS/233/21, and Multi-Scale Medical Robotics Center, InnoHK), in part by Research Grants Council (RGC) of Hong Kong (CUHK 24201219 and CUHK 14217822), in part by The Chinese University of Hong Kong (CUHK) Direct Grant 2021/2022, and in part by project BME-p7-20 of Shun Hing Institute of Advanced Engineering, CUHK. The content is solely the responsibility of the authors and does not necessarily represent the official views of the sponsors.}

\author[MAE,CURI]{Xin Tong}
\ead{xtong@cuhk.edu.hk},
\author[MAE,CURI]{Qingpeng Ding}
\ead{qp.ding@link.cuhk.edu.hk},
\author[MAE,CURI]{Haiyang Fang}
\ead{fyfang@mae.cuhk.edu.hk},
\author[MAE,CURI,MRC,SHI]{Shing Shin Cheng}
\ead{sscheng@cuhk.edu.hk}

\address[MAE]{Department of Mechanical and Automation Engineering, The Chinese University of Hong Kong, Hong Kong}
\address[CURI]{CUHK T Stone Robotics Institute, The Chinese University of Hong Kong, Hong Kong}
\address[MRC]{Multi-Scale Medical Robotics Center, The Chinese University of Hong Kong, Hong Kong}
\address[SHI]{Shun Hing Institute of Advanced Engineering, The Chinese University of Hong Kong, Hong Kong}      

\begin{abstract}
    This paper investigates the robust global attitude stabilization problem for a rigid-body system using quaternion-based feedback. We propose a novel synergistic hybrid feedback with the following notable features: (1) It demonstrates central synergism by utilizing a minimal number of potential functions; (2) It ensures consistency with respect to the unit quaternion representation of rigid-body attitude; (3) Its state-feedback laws incorporate a shared action term that steers the system toward the desired attitude. We demonstrate that the proposed hybrid feedback method effectively solves the problem at hand and guarantees robust uniform global asymptotic stability.
\end{abstract}
  

\end{frontmatter}

\section{Introduction}

Rigid-body attitude stabilization is a fundamental nonlinear control problem with applications in  marine vehicles~\cite{Basso2023}, spacecraft maneuvering~\cite{Shao2023}, and satellite control~\cite{Barman2023}. The rigid-body motion space, known as 3-dimensional special orthogonal group $\SO$, is not contractible, making it impossible to achieve robust global asymptotic stability (GAS) of a particular attitude by continuous state-feedback law~\cite{Bhat2000}. For instance, in~\cite{Maithripala2006,Akhtar2021,Tejaswi2023}, control laws are derived from the gradient of a potential function (PF) on $\SO$ and achieve at most \emph{almost} GAS, where the region of attraction excludes only a set of zero measure. 

Unit quaternions provide a globally nonsingular representation of rigid-body attitude, with each rotation matrix of $\SO$ corresponding to exactly two unit quaternions. Compared to rotation matrices, unit quaternions require fewer parameters, adhere to a simpler composition rule, and are subject to fewer and simpler constraints~\cite{Shuster1993}. Consequently, they are extensively used in attitude control algorithms. However, quaternion-based attitude stabilization requires addressing the following considerations. (i) Control laws have to stabilize the disconnected set of quaternions representing the desired attitude, otherwise \emph{unwinding} may arise, namely yielding an unnecessary full rotation~\cite{Bhat2000,Mayhew2013}. (ii) It is impossible to achieve robust GAS using continuous or memoryless discontinuous quaternion-based state feedback laws~\cite{Sanfelice2006}. (iii) Control laws should exhibit \emph{consistency}, meaning they assign identical values to quaternions representing the same attitude.

\emph{Methodology.} In the past decade, synergistic hybrid feedback has been a promising approach for robust global attitude stabilization. The initial prototype was introduced in quaternion-based hybrid feedback~\cite{Mayhew2011}, followed by a formal proposal on rotation-matrix-based hybrid feedback~\cite{Mayhew2013a}. The key ingredient of this approach is synergistic potential functions (SPFs), i.e., a family of PFs possessing the synergism property: at every critical point (except the desired attitude) of each PF in the family, there exists another PF with a lower value. The gradients of the SPFs constitute a family of state-feedback laws and the synergism property enables a hysteresis-based switching mechanism for state-feedback law selection, ensuring robust GAS results. Hysteresis is used to avoid multiple switches in a short time, which may cause chattering. Specifically, in~\cite{Mayhew2011}, two opposite state-feedback laws---each of which stabilize only one of the two desired quaternions---are selectively employed to rotate the system towards the nearest desired quaternion. In~\cite{Mayhew2013a,Lee2015}, one state-feedback law that is used for stabilization of the desired attitude while other state-feedback laws are employed to steer the system into the domain of attraction of the stabilizing law. These hybrid feedbacks are classified as \emph{noncentrally synergistic}, indicating that they comprises at least one state-feedback law that does not stabilize the desired attitude. If the hysteresis exceeds the upper bound allowed by noncentral synergism, the desired attitude may become unstable. This motivates the development of \emph{centrally synergistic} hybrid feedback in~\cite{Mayhew2011b,Berkane2017a,Berkane2017b,Casau2015,Tong2023}, where each state-feedback law independently stabilizes the system. In other words, regardless of the hysteresis magnitude used, the centrally synergistic approach guarantees at least (local) asymptotic stability.

\emph{Motivation.} Several concerns arise regarding the existing quaternion-based synergistic hybrid feedbacks for attitude stabilization. Firstly, many of these methods, as seen in~\cite{Gui2018,Huang2021,Invernizzi2022,Schlanbusch2015,Zhang2022,Espindola2023}, are adaptations of the work in~\cite{Mayhew2011}, thereby lacking central synergism. Furthermore, they fail to maintain consistency with respect to the quaternion representation of rigid-body attitude. As a result, an additional conversion mechanism is required to properly convert attitude measurements into their quaternion representation. Failure to do so can lead to non-robust stability and unwinding~\cite{Mayhew2013}. Secondly, although the authors in~\cite{Tong2023a} employed the angular warping technique to develop consistent quaternion-based synergistic hybrid feedback, the proposed SPFs consists of six PFs. As a result, this central synergistic control algorithm exhibits significantly larger computational complexity compared to the noncentral synergistic control algorithm in~\cite{Mayhew2011}, even when the latter incorporates the quaternion conversion algorithm in~\cite{Mayhew2013}. Lastly, it is worth highlighting that the robustness achieved in~\cite{Mayhew2011,Tong2023a} is characterized by robust uniform global pre-asymptotic stability. This means that the perturbations allowed by such robustness may result in the closed-loop maximal solution with a bounded time domain. However, it should be noted that such perturbations may be uncommon or improbable in practical scenarios, which diminishes the practical significance of the robustness.

\emph{Contribution.} The main contribution of our work are twofold. Firstly, we further develop the angular warping technique proposed in~\cite{Tong2023a} to propose a consistent centrally synergistic hybrid feedback using a minimal amount of PFs. This design simplifies the switching mechanism compared to~\cite{Tong2023a}. Additionally, unlike~\cite{Mayhew2011}, our method not only eliminates the need for quaternion conversion algorithm but also exhibits a moderate switch of the state-feedback laws, thereby reducing chattering induced by significant measurement noises. Secondly, it is shown that the proposed hybrid feedback ensures robust uniform global asymptotic stability for attitude stabilization. Unlike~\cite{Mayhew2011,Tong2023a}, this robustness result allows maximal solutions to the perturbed closed-loop system to evolve for arbitrarily long hybrid time, thereby being effective against realistic measurement noises and disturbances.

The remainder of this paper is organized as follows. Section~\ref{sec:Preliminaries} introduces preliminary materials and formulates the problem of attitude stabilization. In Section~\ref{sec:SHF}, we present the quaternion-based synergistic hybrid feedback formulation. The main results, including the construction of SPFs and the hybrid controller, are derived in Section~\ref{sec:main}. Simulation results validating the effectiveness of our method are shown in Section~\ref{sec:sim}. Finally, Section~\ref{sec:Conclusion} concludes the paper with closing remarks.

\section{Preliminaries and Problem Formulation} \label{sec:Preliminaries}
In this section, we introduce mathematical notations, offer a brief overview of rigid-body attitude and hybrid dynamical systems, and present the problem formulation.

\subsection{Notations}
We denote by $\R_{\geq 0}$ and $\Z$, the sets of nonnegative real numbers and integer numbers, respectively. For some set $\mc{Z} \subset \mc{X} \times \mc{Y}$, its projection onto $\mc{X}$ is defined as $\Proj_{\mc{X}} \mc{Z} \coloneqq \{x \in \mc{X}: \exists y \in \mc{Y},\; (x,y) \in \mc{Z}\}$; for some set $\mc{U} \subset \mc{X}$, the set $\Proj_{\mc{X}}^{-1} \mc{U}$ is precisely the set $\mc{U} \times \mc{Y}$. The standard Euclidean norm of a vector $x \in \R^n$ is $\Abs{x} \coloneqq \sqrt{x^\top x}$. Let $\mb{B} $ denote the closed unit ball of appropriate dimension in the Euclidean norm. The $n$-dimensional sphere, denoted $\Sphere^n$, is defined as $\Sphere^n \coloneqq \{x \in \R^{n+1}: \Abs{x} = 1\}$. Let $\Pi: \Sphere^n \to \R^{(n+1) \times (n+1)}$ denote the orthogonal projector by $\Pi (x) \coloneqq I_{n+1} - x x^\top$, which projects a vector onto the orthogonal compliment of the unit vector $x$.

\subsection{Attitude Representation by Unit Quaternions}

A rigid-body attitude can be represented by two antipodal points on $\Sphere^3$, which is called unit quaternion and denoted by $Q \coloneqq [\eta , \epsilon^\top]^\top \in \mathbb{S}^3$ with a scalar part $\eta \in \mathbb{R}$ and a vector part $\epsilon \in \mathbb{R}^3$. The identity quaternion is $\idQ \coloneqq [1,0,0,0]^\top$. Let $\times$ denote the cross product on $\R^3$ and define the operator $(\cdot)^\times : \R^3 \to \R^{3 \times 3}$ such that $u^\times v = u \times v$ for all $u,v \in \R^3$. Given a rotational axis $u \in \Sphere^2$ and an angle $\theta \in \R$, we define the two-valued map $\mathcal{Q} : [0,\pi] \times \Sphere^2 \rightrightarrows \Sphere^3$ as $\mathcal{Q} (\theta, u) \coloneqq \left\{ \pm [\cos (\frac{\theta}{2}), \sin (\frac{\theta}{2}) u^\top]^\top \right\}$. The quaternion multiplication is defined as $Q_1 \odot Q_2 \coloneqq [\eta_1\eta_2 - \epsilon_1^\top \epsilon_2, \eta_1 \epsilon_2^\top + \eta_2 \epsilon_1^\top + (\epsilon_1 \times \epsilon_2)^\top]^\top$. The class $\mc{C}^1 (\Sphere^3, \R_{\geq 0})$ denotes the set of nonnegative, continuously differentiable functions on $\Sphere^3$. Given a nonempty set $\indexSet \subset \mb{Z}$, the class $\mc{C}^1_1 (\Sphere^3 \times \indexSet, \R_{\geq 0})$ denotes the set of nonnegative functions on $\Sphere^3$ that are continuously differentiable with respect to their first argument. A function $g: \Sphere^3 \to \R^n$ is said to be \emph{consistent} (with respect to the quaternion representation of rigid-body attitude) if $g(Q) = g(-Q)$ for all $Q \in \Sphere^3$. Define the function $\nu : \R^3 \to \R^4$ by $\nu (\omega) = [0 , \omega^\top]^\top$ and the function $\Lambda : \Sphere^3 \to \R^{4 \times 3}$ by $\Lambda (Q) = [-\epsilon, \eta I_3 - \epsilon^\times]^\top$. 
\begin{lem}\label{lem:Lambda_property}
    The function $\Lambda $ has the following properties:
    \begin{enumerate}[1)]
        \item $\Lambda(Q)^\top \Lambda (Q) = I_3$;
        \item $ \Lambda(Q) \Lambda (Q)^\top  = \Pi (Q)$;
        \item $\Pi (Q) \Lambda(Q) = \Lambda(Q)$;
        \item The matrices $\Lambda (Q)^\top$ and $\Pi (Q)$ have the same null space, i.e., $\nullSpace \Lambda (Q)^\top = \nullSpace \Pi (Q)$.
    \end{enumerate}
\end{lem}
\begin{pf}
    Items 1), 2) have been shown in~\cite{Shuster1993}. Using these two identities, we have that $\Pi (Q) \Lambda(Q) = \Lambda(Q) \Lambda (Q)^\top \Lambda(Q)= \Lambda(Q) $, which verifies item 3). Finally, from items 2) and 3), we can obtain $\nullSpace \Lambda (Q)^\top \subset \nullSpace \Pi (Q)$ and $\nullSpace \Lambda (Q)^\top \supset \nullSpace \Pi (Q)$, respectively, and item 4) is proven.
\end{pf}
The rigid body satisfies the following kinematic and dynamic equations:
\begin{subequations}\label{eq:plant_sys}
    \begin{align}
        \dot{Q} &= \frac{1}{2} Q \odot \nu (\omega) = \frac{1}{2} \Lambda (Q) \omega, \label{eq:dyn_quat} \\
        J \dot{\omega} &= - \omega^\times J \omega + \tau_c   , \label{eq:dyn_angVel}
    \end{align}
\end{subequations}
where $\omega \in \R^3$ denotes the angular velocity expressed in the body-fixed frame, $J \in \R^{3 \times 3}$ is symmetric positive definite and denotes the inertia matrix, and $\tau_c \in \R^3$ denotes the external torque.

\subsection{Hybrid Dynamical Systems}
We refer to the hybrid dynamical system framework proposed in~\cite{Goebel2012,Sanfelice2021}. A hybrid dynamical system with the state $x \in \mathbb{R}^n$, denoted $\mathcal{H}$, can be represented by  
\begin{equation} \label{eq:HySys}
  \mathcal{H} :
  \begin{cases}
    \dot{x} \in F(x) & x \in \mathcal{F},\\
    x^+ \in G(x) & x\in \mathcal{J},
  \end{cases}
\end{equation}
where the set-valued map $F: \mathbb{R}^n \rightrightarrows \mathbb{R}^n$ is the \emph{flow map} capturing the continuous evolution on the \emph{flow set} $\mathcal{F} \subset \mathbb{R}^n$, the set-valued map $G: \mathbb{R}^n \rightrightarrows \mathbb{R}^n$ is the \emph{jump map} capturing the discrete evolution on the \emph{jump set} $\mathcal{J} \subset \mathbb{R}^n$, and $x^+$ indicates the values of the state after the jump. The \emph{hybrid time domain} is a subset $E \subset \R_{\geq} \times \mathbb{N}$ such that $E = \bigcup_{j=0}^{J-1}([t_j,t_{j+1}], j)$ for some finite sequence of times $0 = t_0 \leq t_1 \leq \dots \leq t_J$. A function $\phi: \domain \phi \to \R^n$ is said to be a \emph{solution} to system $\mc{H}$ if $\domain \phi$ is a hybrid time domain and $\phi$ satisfies the dynamics and constraints given by~\eqref{eq:HySys}. A solution $\phi$ to $\mc{H}$ is said to be: \emph{complete} if $\domain \phi$ is unbounded; \emph{maximal} if it cannot be extended to another solution; \emph{precompact} if it is complete and its range is bounded; \emph{eventually continuous} if $J = \sup_{j} \domain \phi < \infty$ and $\domain \phi \bigcap (\R_{\geq 0} \times \{J\})$ contains at least two points. The following stability definition will be used throughout this paper.

\begin{defn}
    Let $\SetPoint \subset \R^n$ be compact. The distance of a point $x \in \R^n$ to $\SetPoint$, denoted $\Abs{x}_{\SetPoint}$, is defined by $\Abs{x}_{\SetPoint}\coloneqq \inf_{y \in \SetPoint}\Abs{x - y}$. For system $\mc{H}$ in~\eqref{eq:HySys}, the compact set $\mc{A}$ is said to be:
    \begin{enumerate}[1)]
        \item \emph{uniformly globally pre-asymptotically stable} (\emph{UGpAS}) if there exists a function $\beta \in \mc{KL} $ such that each solution $\mc{\phi}$ to $\mc{H}$ satisfies
        \begin{align*}
            \Abs{\phi(t,j)}_{\SetPoint} &\leq \beta (\Abs{\phi(0,0)}_{\SetPoint}, t+j) & \forall (t,j) \in \domain \phi ;
        \end{align*}
        \item \emph{robustly UGpAS} if there exists a continuous function $\rho: \R^n \to \R_{\geq 0}$ that is positive on $\R^n \setminus \SetPoint$ such that $\SetPoint$ is UGpAS for $\mc{H}_{\rho}$, the $\rho$-perturbation of $\mc{H}$, which is defined as 
        \begin{align}
            \mc{H}_{\rho} : 
            \begin{cases}
                \dot{x} \in F_{\rho}(x) & x \in \mathcal{F}_{\rho},\\ 
                x^+ \in G_{\rho}(x) & x\in \mathcal{J}_{\rho},
            \end{cases} \label{eq:perturb_HySys}
        \end{align}
        where $\mc{F}_{\rho} = \{x \in \R^n: (x+\rho (x) \mb{B}) \bigcap \mc{F} \neq \emptyset\}$, $F_{\rho} (x) = \overline{\operatorname{con}} F ((x+\rho (x) \mb{B}) \bigcap \mc{F}) +  \rho (x) \mb{B}$\footnote{Given a set $A$, $\overline{\operatorname{con}} A$ denote the closed convex hull of the set $A$.}, $\mc{J}_{\rho} = \{x \in \R^n: (x+\rho (x) \mb{B}) \bigcap \mc{J} \neq \emptyset\}$, and $G_{\rho} (x) = \{v \in \R^n : v \in g + \rho(g) \mb{B}, g \in G((x+\rho (x) \mb{B}) \bigcap \mc{J})\}$.
    \end{enumerate}
    The prefix ``pre'' means that maximal solutions need not be complete, and it is dropped when every maximal solution is complete.
\end{defn}

\subsection{Problem Formulation}

Let $\SetPoint \coloneqq \{\idQ, -\idQ\} \subset \Sphere^3 $ denote the desired attitude, and define the compact set $\mc{O}_p \coloneqq \{(Q,\omega) \in \Sphere^3 \times \R^3: Q \in \SetPoint, \omega = 0\}$. The robust global attitude stabilization problems are formulated as follows.

\begin{prob}\label{prob:GAS_sub}
    Given system~\eqref{eq:dyn_quat} with control variable $\omega$, design a state-feedback controller such that the set $\mc{A}  $ is robustly UGAS.
\end{prob}

\begin{prob}\label{prob:GAS}
    Given system~\eqref{eq:plant_sys} with control variable $\tau_c$, design a state-feedback controller such that the set $\mc{O}_p  $ is robustly UGAS.
\end{prob}

\begin{rem}
    The unwinding phenomenon pertains to a behavior of the kinematics system~\eqref{eq:dyn_quat}, where the system starts from an attitude arbitrarily close to the desired attitude and undergoes a significant rotation before reaching the desired attitude. Consequently, this unwinding behavior cannot occur when the desired attitude is stable for system~\eqref{eq:dyn_quat}. In this sense, we assert that solutions to Problems~\ref{prob:GAS_sub},~\ref{prob:GAS} can effectively avoid unwinding.
\end{rem}

\section{Synergistic Hybrid Feedback} \label{sec:SHF}
In this section, we begin by introducing the definition of the SPFs for quaternion-based attitude stabilization. Then, we derive the hybrid feedback by utilizing the gradients of the SPFs and establish its effectiveness for Problem~\ref{prob:GAS_sub}. Furthermore, we examine existing methods within this framework, offering insights that motivate our subsequent results in the next section.

\subsection{Synergistic Potential Functions}
\begin{defn}\label{defn:PF}
    A function $V \in \mc{C}^1 (\Sphere^3, \R_{\geq 0})$ is called the \emph{potential function relative to the set $\SetPoint$} if $ V $ is positive definite relative to $\SetPoint $.
\end{defn}
Let the Riemannian metric on $\Sphere^3$ be induced by the Euclidean inner product on $\R^4$. Given a potential function $V$ relative to $\SetPoint$, we denote by $\gradA V $ and $\grad V$, the gradient of $V$ on $\R^4$ and $\Sphere^3$, respectively. That is, $\gradA V (Q) \coloneqq (\frac{\partial V(Q)}{\partial Q} )^\top$ is the column vector of partial derivatives, and $\grad V (Q) \coloneqq \Pi(Q) \gradA V (Q)$. The set of \emph{critical points} of $V$, denoted $\CritPoint V$, is defined as $\CritPoint V \coloneqq \{ Q \in \Sphere^3 : \grad V (Q) = 0\}$. From the Definition~\ref{defn:PF}, it follows that $\SetPoint \subset \CritPoint V$. 
 
\begin{exmp}\label{exmp:basic_PF}
    Consider the potential function $P \in \mc{C}^1 (\Sphere^3 , \R_{\geq 0})$, which is defined as 
    \begin{align}
        P(Q) = \epsilon^\top A \epsilon, \label{eq:basic_PF}
    \end{align} 
    where $A = A^\top \in \R^{3 \times 3}$ is positive definite. The set of critical points of $P$ is given by 
    \begin{align}
        \CritPoint P = \SetPoint \bigcup \left\{\mathcal{Q} (\pi, v  ) \in \Sphere^2: v \in \mathcal{E} (A) \right\}
    \end{align}
    where $\mathcal{E} (A)$ denote the set of real unit eigenvectors of $A$. The function $P$ and its variants are commonly used in the attitude control~\cite{Mayhew2013a,Berkane2017a,Berkane2017b,Tong2023a}.
\end{exmp}

Let $\indexSet \subset \mathbb{Z}$ be a finite index set. A function $U  \in \CPF $ can be regarded as a family of potential functions on $\Sphere^3$, which is indexed by the logic variable $q$. The \emph{synergy gap} of $U$ is defined as $\mu_U( Q,q) \coloneqq U(Q,q) - \min_{p\in \indexSet} U(Q,p)$. Let $\gradA_1 {U}$ and $\grad_1 U$ denote the gradient of $U$ with respect to its first argument on $\R^4$ and $\Sphere^3$, respectively. That is, $\gradA_1 U(Q,q) \coloneqq (\frac{\partial U(Q,q)}{\partial Q} )^\top$ is the column vector of partial derivatives with respect to the first argument, and $\grad_1 U (Q,q) \coloneqq \Pi(Q) \gradA_1 U(Q,q)$. The set of \emph{critical points} of $U$ is defined as $ \CritPoint U \coloneqq \{(Q,q) \in \Sphere^3 \times \indexSet :  \grad_1 {U} (Q,q)= 0 \}$.

\begin{defn}[\cite{Mayhew2013a,Sanfelice2021}] \label{defn:syn_defn}
    A function $U \in \CPF $ is said to be \emph{synergistic relative to $\SetPoint$ with gap exceeding $\delta$} if $\delta > 0$ and there exist a set $\SetObject  \subset \SetPoint \times \indexSet$ such that the following hold:
    \begin{enumerate}[P1)]
        \item $\Proj_{\Sphere^3} \SetObject = \SetPoint$;
        \item $U$ is positive definite relative to $\SetObject$;
        \item $\mu_U (Q,q) > \delta$ for all $(Q,q) \in (\CritPoint U \bigcup (\SetPoint \times \indexSet) ) \setminus \SetObject$.
    \end{enumerate}  
    The synergism property is said to be \emph{central} if $\SetObject = \SetPoint \times \indexSet$, and \emph{noncentral} otherwise. We shall call $U$ the synergistic potential functions (SPFs).
\end{defn}

\subsection{Hybrid Feedback}

The gradients of the SPFs $U$ satisfying Definition~\ref{defn:syn_defn} induces a family of state feedback laws $\kappa_U : \Sphere^3 \times \indexSet \to \R^3$ for system~\eqref{eq:dyn_quat}, which is defined as 
\begin{align}
    \kappa_U (Q,q) \coloneqq \Lambda^\top (Q) \grad_1 U (Q,q) = \Lambda^\top (Q) \gradA_1 U(Q,q), \label{eq:hyrid_feedback}
\end{align} 
where the second equality follows from item 3) of Lemma~\ref{lem:Lambda_property}. Further, the next result follows immediately from item 4) of Lemma~\ref{lem:Lambda_property}:
\begin{lem} \label{lem:kappa_crit}
    Given any function $U \in \CPF$, $\kappa_U (Q,q) = 0$ if and only if $(Q,q) \in \CritPoint U$.
\end{lem}

By~\eqref{eq:hyrid_feedback}, we synthesize the synergistic hybrid controller for system~\eqref{eq:dyn_quat}:
\begin{subequations} \label{eq:ctr_sys}
    \begin{align}
        & \omega = - k_p \kappa_U (Q,q)  , \\
        & 
        \begin{cases}
            \dot{q} = 0 & (Q,q) \in \mathcal{F}_K , \\
            q^+ \in \argmin_{p \in \indexSet} U(Q,p) & (Q,q) \in \mathcal{J}_K,
        \end{cases} \label{eq:ctr_dyn}
    \end{align}
\end{subequations}
where $k_p$ is positive gain, the flow set and jump set are defined as $\mathcal{F}_K \coloneqq \{(Q,q) \in \Sphere^3 \times \indexSet : \mu_U (Q,q) \leq \delta_h\}$ and $\mathcal{J}_K \coloneqq \{(Q,q) \in \Sphere^3 \times \indexSet : \mu_U (Q,q) \geq \delta_h\}$, respectively, and the constant $\delta_h$ fulfills $0 < \delta_h \leq \delta$. 

The controller~\eqref{eq:ctr_sys} retains a logic state $q \in \indexSet$, which is governed by the hybrid dynamics in~\eqref{eq:ctr_dyn}. Indeed, \eqref{eq:ctr_dyn} can be regarded as a hysteresis-based \emph{switching mechanism}. Specifically, the state $q$ is hysteretically switched to $\argmin_{p \in \indexSet} U(Q,p)$, so as to preclude the state-feedback law $\kappa_U(Q,q)$ from vanishing when $Q$ approaches the desired attitude. The scalar $\delta_h$ denotes the \emph{hysteresis width}, quantifying the extent of hysteresis. The state-feedback laws~\eqref{eq:hyrid_feedback} in conjunction with the switching mechanism~\eqref{eq:ctr_dyn} is so-called \emph{hybrid feedback}.

\subsection{Closed-loop Stability}

Let $\mathcal{X}_1 \coloneqq \Sphere^3   \times \indexSet $ and $x_1 \coloneqq (Q, q)$ denote the state space and state, respectively. Applying controller~\eqref{eq:ctr_sys} to system~\eqref{eq:dyn_quat} yields the closed-loop system as follows:
\begin{align}
    \mathcal{H}_1 :& 
    \begin{cases}
        \dot{x}_1 = F_1 (x_1) & x_1 \in \mathcal{F}_1 ,\\
        x_1^+ \in G_1 (x_1) & x_1 \in \mathcal{J}_1,
    \end{cases} \label{eq:clp_sys}
\end{align}
where the flow set and jump set are defined as $ \mathcal{F}_1 \coloneqq \mc{F}_K$ and $\mc{J}_1 \coloneqq \mc{J}_K$, respectively, and the flow map and jump map are defined as
\begin{align*}
    F_1(x_1) &:
    \begin{cases}
        \dot{Q} = -\frac{k_p}{2} \grad_1 U(Q,q), \\ 
        \dot{q} = 0,
    \end{cases}
    \\
    G_1(x_1) & \coloneqq \left(Q,  \argmin_{p \in \indexSet} U(Q,p) \right) ,
\end{align*}
where we make use of item 2) of Lemma~\ref{lem:Lambda_property} and the fact that $\Pi(Q) \grad_1 U(Q,q) = \grad_1 U(Q,q)$ for all $(Q,q) \in \Sphere^3 \times \indexSet$.

Define the compact set $ \mc{O}_1 \coloneqq \SetPoint \times \indexSet \subset \mc{X}_1 $. The sets $\mc{A}$, $\mc{B}$, and $\mc{O}_1$ satisfy the relations: $\mc{B}  \subset \mc{O}_1$, $ \Proj_{\Sphere^3   } \mc{O}_1 = \mc{A}$, $\Proj_{\Sphere^3 \times \indexSet }^{-1} \mc{A}  = \mc{O}_1$. The last relation implies that the stability property can be passed from $\mc{O}_1$ to $\mc{A}$. Let us make this precise. 

\begin{lem}\label{lem:stability_pass}
    If the set $\mc{O}_1$ is (robustly) UGAS for system $\mc{H}_1$ given by~\eqref{eq:clp_sys}, then the set $\mc{A}$ is (robustly) UGAS for its subsystem~\eqref{eq:dyn_quat}.
\end{lem}
\begin{pf}
    Let $\phi$ be a solution to $\mc{H}_1$, and let $\phi_a$ denote the coordinate of $\phi$ on the subsystem~\eqref{eq:dyn_quat}. Since the stability property of a set is characterized by the distance from the set and since $\Abs{\phi (t,j)}_{\mc{O}_1} = \Abs{\phi_a (t,j)}_{\mc{A}}$, the lemma follows.
\end{pf}

Now we are ready to certify the solvability of Problem~\ref{prob:GAS_sub} by controller~\eqref{eq:ctr_sys}.
\begin{thm}\label{thm:stability}
    Problem~\ref{prob:GAS_sub} is solvable by controller~\eqref{eq:ctr_sys} in the sense that the set $  \mc{O}_1 $ is robustly UGAS for system $\mc{H}_1$ given by~\eqref{eq:clp_sys}.
\end{thm}
\begin{extendVersion}
\begin{pf}
    The proof entails verifying statements (S1) through (S6) as follows.

    \emph{(S1) The autonomous system $\mc{H}_1$ satisfies the hybrid basic conditions.} 
    
    It suffices to show that \eqref{eq:clp_sys} satisfies items (A1)-(A3) of Definition~\ref{defn:basic_Hy_condition}. First, $\mc{F}_1$ and $\mc{J}_1$ are closed thanks to the continuity of $\mu_U$, so item (A1). Second, $F_1$ is single-valued and continuous, so item (A2) is satisfied. Finally, $G_1 $ is locally bounded since $\mc{X}_1$ is compact. In addition, for any sequence $z_i \in \mc{X}_1$ such that $z_i \to z $ as $ i \to \infty $, the outer limit of of $G_1(z_i)$ equals the closed set $G_1 (z)$. Hence, $G_1$ is outer semicontinuous, which shows item (A3). 

    \emph{(S2) Each maximal solution to $\mc{H}_1$ is precompact and eventually continuous.} 
    
    Consider a Lyapunov function candidate $V_1 : \mc{X}_1 \to \R_{\geq 0}$ defined as $V_1 (x_1) \coloneqq U(Q,q)$. By item P1) of Definition~\ref{defn:syn_defn}, $V_1$ is positive definite relative to $\mc{B}$. The change of $V_1$ at each $x_1 \in \mc{F}_1$ along flows of~\eqref{eq:clp_sys} is given by 
    \begin{align}
        \dot{V}_1(x_1) 
        & =  - \frac{k_p}{2}  \Abs*{\grad_1 U(Q,q)}^2  \leq 0 . \label{eq:flow_V_1}
    \end{align}
    For each $x_1\in \mc{J}_1$, the change of $V_1$ over jumps of~\eqref{eq:clp_sys} is given by 
    \begin{align}
        \Delta V_1(x_1) &= \max_{z \in G_1 (x_1)}{V_1 (z)} - V_1(x_1) \notag \\ 
        & = -  \mu_U(Q,q) \leq -\delta_h < 0. \label{eq:jump_V_1}
    \end{align}
    Therefore, $V_1$ is nonincreasing along solutions to $\mc{H}_1$. Since $V_1$ is proper, it follows that solutions to $\mc{H}_1$ are bounded and have finitely many jumps. In addition, we make the following observations: For each $ x_1 \in \mc{F}_1 \setminus \mc{J}_1 = \{x_1 \in \mc{X}_1: \mu_{U} (Q,q) < \delta_h\}$, $F_1(x_1)$ intersects with the tangent cone to $\mc{F}_1$ at $x_1$, which is due to the openness of $\mc{F}_1 \setminus \mc{J}_1$; $G_1(\mc{J}_1) \subset \mc{F}_1 \setminus \mc{J}_1 \subset \mc{X}_1$. Combining these observations, we have established the statement (S2) by invoking Proposition~\ref{prop:existence_condition}.

    \emph{(S3) The set $\mc{B}$ is UGAS for $\mc{H}_1$.}
    
    According to P2), P3) of Definition~\ref{defn:syn_defn} and $\delta_h \leq \delta$, we have that the flow set $\mc{F}_1 $ contains the set $\SetObject$, but does not intersect the set $(\CritPoint U \bigcup (\SetPoint \times \indexSet) ) \setminus \SetObject$, thereby $\mc{F}_1 \bigcap\CritPoint U =  \SetObject$. It follows from~\eqref{eq:flow_V_1} that $V_1$ is strictly deceasing for all $x_1 \in \mc{F}_1 \setminus \SetObject$. By invoking hybrid Lyapunov theorem~\cite[Theorems 3.19 \& 3.22]{Sanfelice2021} and statement (S2), we can arrive at statement (S3).

    \emph{(S4) The set $\mc{O}_1$ is UGAS for $\mc{H}_1$.}
    
    According to~\cite[Propositin 7.5 \& Theorem 7.12]{Goebel2012} and $\mc{B} \subset \mc{O}_1$, it suffices to verify that $\mc{O}_1$ is forward invariant. For each maximal solution $\phi$ to $\mc{H}_1$, we make the following observation: If $\phi(0,0) \in \mc{B} \subset \mc{F}_1 \setminus \mc{J}_1$, $\phi$ remains unchanged in $\mc{B}$, since $F_1$ vanishes on $\mc{B}$; If $\phi(0,0) \in \mc{O}_1 \setminus \mc{B} \subset \mc{J}_1 \setminus \mc{F}_1$, $\phi$ takes a single immediate jump into the set $\mc{B}$, which is due to $G_1 (\mc{O}_1) = \mc{B}$. Therefore, $\mc{O}_1$ is forward invariant, as to prove.

    \emph{(S5) The set $\mc{O}_1$ is robustly UGAS for $\mc{H}_1$.} 
    
    From statements (S1), (S4) and~\cite[Theorem 7.21]{Goebel2012}, we assert that the set $\mc{O}_1$ is UGpAS for $\mc{H}_{1,\rho}$, the $\rho$-perturbation of $\mc{H}_1$, where $\rho : \mc{X}_1 \to \R_{\geq 0}$ is a continuous function that is positive on $\mc{X}_1 \setminus \mc{O}_1 $. By~\cite[Proposition 6.28]{Goebel2012}, $\mc{H}_{1,\rho}$ satisfies the hybrid basic conditions. From~\eqref{eq:perturb_HySys}, we have that $\mc{F}_1 \subset \mc{F}_{1,\rho} \subset \mc{X}_1$ and $\mc{J}_1 \subset \mc{J}_{1,\rho} \subset \mc{X}_1$, which yields $\mc{F}_{1,\rho} \bigcup \mc{J}_{1,\rho} = \mc{X}_1$ due to $\mc{X}_1 = \mc{F}_1 \bigcup \mc{J}_1 $. Combining these set relations yields 
    $\mc{F}_{1,\rho} \setminus \mc{J}_{1,\rho} = \mc{F}_{1,\rho} \bigcap (\mc{X}_1 \setminus \mc{J}_{1,\rho}) =  \mc{F}_{1,\rho}  \bigcap ( (\mc{F}_1 \bigcup \mc{J}_1) \setminus \mc{J}_{1,\rho})  = \mc{F}_{1,\rho}  \bigcap ( (\mc{F}_1 \setminus \mc{J}_1) \setminus \mc{J}_{1,\rho})  = (\mc{F}_1 \setminus \mc{J}_1) \setminus \mc{J}_{1,\rho} $. Since the set $\mc{F}_1 \setminus \mc{J}_1$ is open and the set $\mc{J}_{1,\rho}$ is closed, it follows that the set $\mc{F}_{1,\rho} \setminus \mc{J}_{1,\rho}$ is open. Then, we make the following observations: For each $ x_1 \in \mc{F}_{1,\rho} \setminus \mc{J}_{1,\rho} $, $F_{1,\rho}(x_1)$ intersects with the tangent cone to $\mc{F}_{1,\rho}$ at $x_1$; Solutions to $\mc{H}_{1,\rho}$ cannot blow up to infinity at a finite time, due to the UGpAS property of the compact set $\mc{O}_1$; $G_{1,\rho}(\mc{J}_{1,\rho}) \subset \mc{F}_{1,\rho} \bigcup \mc{J}_{1,\rho}$. According to these observations, we can conclude that maximal solutions to $\mc{H}_{1,\rho}$ are complete as per Proposition~\ref{prop:existence_condition}, and statement (S5) follows. This completes the proof. 
\end{pf} 
\end{extendVersion}

\subsection{Discussion}
\begin{figure}[htb]
    \centering
    \includegraphics[width=8cm]{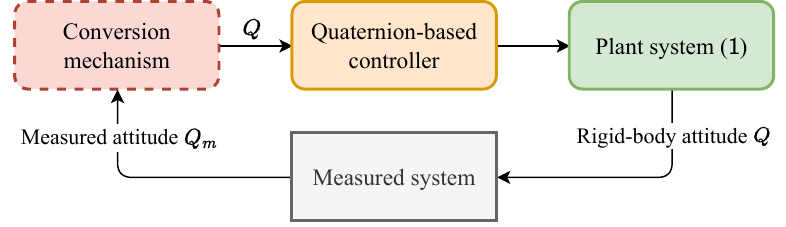}
    \caption{Flowchart of quaternion-based attitude control.}
    \label{fig:flowchart}
\end{figure}

Several comments are in order about controller~\eqref{eq:ctr_sys}:
\begin{enumerate}[(i)]
    \item Hysteresis is necessary for controller~\eqref{eq:ctr_sys} to achieve robustly UGAS result.
    \item Central synergism is generally regarded as more desirable in practice than noncentral synergism.
    \item If the hybrid feedback~\eqref{eq:hyrid_feedback} is inconsistent, the correctness of Theorem~\ref{thm:stability} depends on using an additional mechanism to uniquely select the quaternion measurement that satisfies~\eqref{eq:dyn_quat}, see Fig.~\ref{fig:flowchart}.
    \item Minimal cardinality of the set $\indexSet$ is always advantageous in terms of computation complexity, since the switching mechanism requires to evaluate the SPFs for each $q \in \indexSet$. 
\end{enumerate}

The next result clarifies the comments (i), (ii).
\begin{cor}\label{cor:hysteresis}
    Consider system~\eqref{eq:clp_sys}. The following hold for variant hysteresis width $\delta_h $ in~\eqref{eq:ctr_sys}:
    \begin{enumerate}[1)]
        \item If $\delta_h = 0 $, then $\mc{B}$ is stable, but not UGAS. 
        \item If $\delta_h > \max_{(Q,q) \in \mc{N}} \mu_U (Q,q) $ with $\mc{N} = \overline{\CritPoint U} \bigcup (\SetPoint \times \indexSet) $, then $\mc{B}$ is locally asymptotically stable.
    \end{enumerate}
\end{cor}
\begin{extendVersion}
\begin{pf}
    By similar arguments in the proof of Theorem~\ref{thm:stability}, system $\mc{H}_1$ satisfies the hybrid basic conditions and therefore is well-posed. We continue using the Lyapunov function candidate $V_1 = U$. Since $\delta_h \geq 0$ and since $V_1 $ is proper, we have that $V_1$ is nonincreasing and that each maximal solution to $\mc{H}_1$ is precompact. Therefore, $\mc{B}$ is stable as per~\cite[Theorem 3.19]{Sanfelice2021}.

    If $\delta_h = 0$, it follows from hybrid invariance principal that there exist eventually discrete solutions to $\mc{H}_1$ that remain in the set $\{x_1 \in \mc{X}_1 : \mu_U (Q,q) = 0\} $, and so the set $\mc{B}$ fails to be UGAS, as shown in item 1). 

    If $\delta_h > \max_{(Q,q) \in \mc{N}} \mu_U (Q,q) $, then the flow set $\mc{F}_1$ contains the set $ \mc{M}_1 = \{x_1 \in \mc{X}_1 : (Q,q) \in \CritPoint U \bigcup (\SetPoint \times \indexSet) \setminus \SetObject \}$, which contains the undesired equilibrium points of system $\mc{H}_1$. Therefore, neither $\mc{B}$ nor $\mc{O}_1$ is UGAS. Consider the open set $\mc{U}_{ 1} = \{x_1 \in \mc{X}_1 : V_1(x_1) < \delta/2\}$. From the synergism property of $U$, we conclude that $ \mc{B} \subset \mc{U}_{ 1} \subset \mc{F}_1$ and $\mc{M}_1 \bigcap \mc{U}_{ 1} = \emptyset$. According to hybrid invariance principle, solutions to $\mathcal{H}_1$ with initial conditions in $\mc{U}_1$ converges to $\mc{B}$. This proves item 2).
\end{pf}
\end{extendVersion}

\begin{rem}
    Corollary~\ref{cor:hysteresis} shows that the set $O_1$ fails to be UGAS when no hysteresis is used in controller~\eqref{eq:ctr_sys}. On the other hand, if the hysteresis width is chosen to be excessively large, the controller will only guarantee local asymptotic stability of the set $\mathcal{B}$. If controller~\eqref{eq:ctr_sys} is centrally synergistic, it still results in the asymptotic stability of $\mathcal{O}_1$ due to $\mc{O}_1 = \mc{B}$. Therefore, we can conclude that the centrally synergistic hybrid feedback avoids unwinding \emph{intrinsically} (i.e., without relying on the switching mechanism).
\end{rem}

The following example clarifies the comment (iii).

\begin{figure}[htb]
    \centering
    \includegraphics[width=8cm]{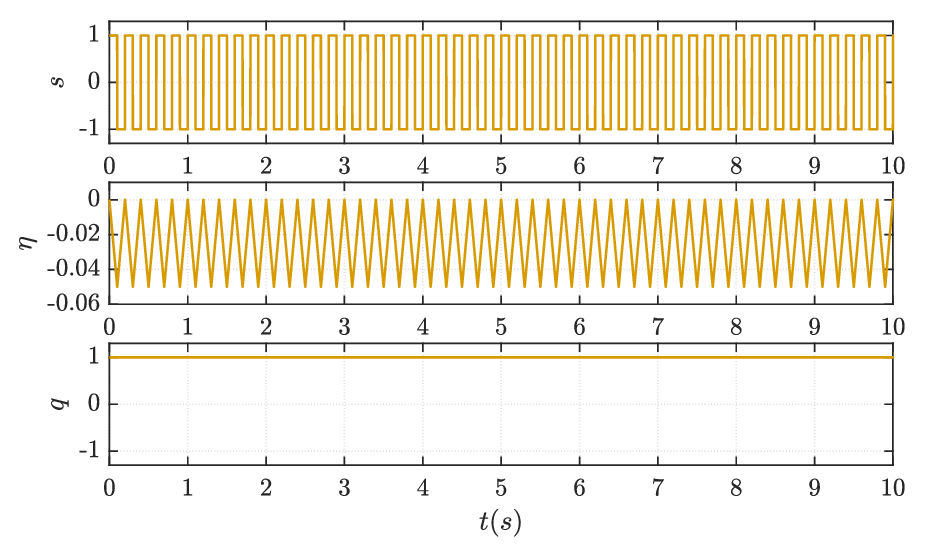}
    \caption{Example~\ref{exmp:non_cen_SPF}: Chattering at a neighborhood of $\eta = 0$ when the hybrid feedback from~\eqref{eq:non_SPF} is applied to Problem~\ref{prob:GAS_sub} by directly using quaternion measurement $Q_m = sQ$, where $s(t)$ is a square wave with a frequency of $5\mathrm{Hz}$. Initial conditions are set as $Q(0) = [0,0.6,0.8,0]^\top$ and $q(0) = 1$.}
    \label{fig:FirstOrderInconsistent_231204}
\end{figure}

\begin{exmp}[\cite{Mayhew2011}] \label{exmp:non_cen_SPF}
    Let $\indexSet = \{-1,1\}$. Define the function $U \in \CPF $ as 
    \begin{align}
        U(Q,q) = 1 - q \eta, \label{eq:non_SPF}
    \end{align}
    which is \emph{noncentrally synergistic} relative to $\SetPoint$ with the parameters $\SetObject = \{(\idQ, 1), (-\idQ, -1)\}$ and $\delta \in (0,1)$. It has the following properties:
    \begin{enumerate}[1)]
        \item The set of critical points of $U$ are $\CritPoint U = \SetPoint \times \indexSet $. 
        \item The feedback~\eqref{eq:hyrid_feedback} is explicitly expressed as $\kappa_U (Q,q) = q \epsilon$ and is \emph{inconsistent} with respect to the unit quaternion for each fixed $q$.
        \item The switching mechanism~\eqref{eq:ctr_dyn} operates on the flow set $\mc{F}_K = \{(Q,q) \in \Sphere^3 \times \indexSet : 2 q \eta \geq -\delta_h \}$ and the jump set $\mc{J}_K = \{(Q,q) \in \Sphere^3 \times \indexSet : 2 q \eta \leq -\delta_h \}$ with $0< \delta_h < 1$.
    \end{enumerate}
    Adversarial selection of quaternion measurements can cause the failure of this hybrid feedback to achieve the UGAS results, as shown in Fig.~\ref{fig:FirstOrderInconsistent_231204}.
\end{exmp}

\section{Main Results} \label{sec:main}
In this section, we begin by introducing the SPFs candidate that consists of two potential functions. Subsequently, the parameters are determined to ensure both the central synergism and consistency properties. Furthermore, the hybrid feedback is derived from the SPFs, which is effectively employed to tackle the problem of stabilizing rigid-body attitude.

\subsection{SPFs Candidate}
Let the index set $\indexSet \coloneqq \{-1,1\}$. Let $u \in \Sphere^2$ be a constant unit vector to be determined and define a pair of vectors $\{u_q \in \Sphere^2:  q \in \indexSet\}$ as $u_q \coloneqq q u$. Define the skew-symmetrical matrix $S_q \coloneqq  \nu (u_q) \idQ^\top - \idQ \nu (u_q)^\top \in \R^{4 \times 4}$ for each $q \in \indexSet$. Let $\theta: \Sphere^3 \to \R_{\geq 0}$ be a potential function relative to $\SetPoint$. By using the parameter $u_q$ and the function $\theta$, the angular warping transformation $\Warp : \Sphere^3 \times \indexSet \to \Sphere^3$ is defined by 
\begin{align}
    \mathcal{T}(Q,q) &\coloneqq e^{S_q \theta (Q)} Q  \notag \\
    &= 
    \begin{bmatrix}
        \cos (\theta(Q)) & - \sin (\theta(Q))  u_q^\top \\
        \sin (\theta(Q))  u_q   & \Pi (u_q) + \cos(\theta(Q)) u_q u_q^\top 
    \end{bmatrix} Q \notag \\
    &= 
    \begin{bmatrix}
        \Xi(Q,q) \\
        \epsilon  +  \Gamma (Q,q)u_q
    \end{bmatrix}   , \label{eq:warp_quat}
\end{align}
where the matrix $e^{S_q \theta (Q)}$ is expanded by using the generalized Rodrigues' formula~\cite{Gallier2003}, and the functions $\Xi, \Gamma : \Sphere^3 \times \indexSet\to \R $ are defined as 
\begin{align}
    \Xi(Q,q) & \coloneqq   \cos(\theta (Q)) \eta - \sin(\theta (Q))  (u_q^\top \epsilon) , \label{eq:def_Xi} \\
    \Gamma (Q,q) & \coloneqq \sin(\theta (Q)) \eta + \left(\cos(\theta (Q)) - 1\right) (u_q^\top \epsilon)  . \label{eq:def_Gamma}
\end{align}
Geometrically, $\Warp$ rotates a unit quaternion $Q \in \Sphere^3$ by an orthonormal matrix $e^{S_q \theta (Q)}$. The rotation angle is described by the value of function $\theta$ at the point $Q$. The rotation direction is described by the matrix $S_q$ and is determined by the constant vector $u_q$.  

We now define the SPFs candidate $U : \Sphere^3 \times \indexSet\to \R_{\geq 0}$ by the composition of $P$ in~\eqref{eq:basic_PF} and $\Warp$ in~\eqref{eq:warp_quat}, i.e., $U \coloneqq P \circ \Warp$, which can be expressed as 
\begin{align}
    & U(Q,q) = \epsilon^\top A \epsilon + 2 \Gamma(Q,q) (u_q^\top A \epsilon ) + \Gamma(Q,q)^2  (u_q^\top A u_q   )  . \label{eq:SPF_defn}
\end{align}

We emphasize that the transformation in~\eqref{eq:SPF_defn} was proposed in the authors' previous work~\cite{Tong2023a}. In that work, the central synergism is achieved using six vectors $u_q$. However, in this paper, our objective is to ensure the central synergism of~\eqref{eq:SPF_defn} with only two vectors $u_q$. It will allow us to minimize the number of potential functions required for the synergism, thereby reducing the computational complexity of the switching mechanism.

\subsection{Central Synergism and Consistency}
To achieve the central synergism of~\eqref{eq:SPF_defn}, we shall need the following assumption.  

\begin{assum}\label{assum:A}
    The parameter matrix $A$ has distinct eigenvalues $\{\lambda_1,\lambda_2,\lambda_3\}$ such that $0 < \lambda_1 < \lambda_2 < \lambda_3$. 
\end{assum}
\begin{rem}
    The matrices that satisfy Assumption~\ref{assum:A} are dense in the set of symmetric positive definite matrices in $\R^{3 \times 3}$. Hence, Assumption~\ref{assum:A} is a mild condition. On the other hand, if the matrix $A$ has repeated eigenvalues, it becomes problematic to obtain the synergism for $U$ in~\eqref{eq:SPF_defn}, which will be clarified in Corollary~\ref{cor:assum}.  
\end{rem}

Let $ k \in (0,\lambda_1 / \lambda_3)$ be a constant scalar. The warping angle $\theta$ is defined as 
\begin{align}
    \theta (Q) &\coloneqq k \epsilon^\top \epsilon . \label{eq:warp_angle}  
\end{align} 

\begin{thm} \label{thm:cspf}
    Consider the SPFs candidate $U$ defined as in~\eqref{eq:SPF_defn} with $\theta$ given by~\eqref{eq:warp_angle} under Assumption~\ref{assum:A}. For the $i$th eigenvalue $\lambda_i$ of $A$, we denote by $v_i$ the corresponding $i$th unit eigenvector of $A$. 
    Then, the following hold:
    \begin{enumerate}[1)]
        \item $U$ and $\kappa_U$ are consistent;
        \item If $u^\top v_i \neq 0$ for each $i \in \{ 1,2,3\}$, then $U$ is centrally synergistic relative to $\SetPoint $ with gap exceeding some positive scalar $\delta $.
        \item If $u = \frac{1}{\sqrt{3}} (v_1 + v_2 + v_3)$, then $ \delta = \frac{4}{3} \sin^2(k - \frac{k^3}{3}  ) \bigl(\lambda_1 - \frac{\lambda_1+\lambda_2+\lambda_3}{3} \sin^2(k) \bigr) $. 
    \end{enumerate} 
\end{thm}
\begin{pf}
    The proof is divided into three parts. 

    \textbf{Part 1.} Since the function $\theta$ given by~\eqref{eq:warp_angle} is consistent, it follows from~\eqref{eq:warp_quat} that $ \Warp(-Q,q) = -e^{S_q \theta(Q)} Q = - \Warp(Q,q) $ for all $(Q,q) \in \Sphere^3 \times \indexSet$. Note that $P$ is also consistent. According to $U = P \circ \Warp$, it follows that $U(Q,q) = U(-Q,q)$ and $\gradA_1 U(Q,q) = - \gradA_1U(-Q,q)$. It implies $\kappa_U (Q,q) = \kappa_U (-Q,q)$, which is due to $\Lambda(Q) = -\Lambda(-Q)$. This shows item 1).

    \textbf{Part 2.} The proof of item 2) entails establishing the statements (S1)-(S3) as follows. Let $\SetObject = \SetPoint \times \indexSet$. 

    \emph{(S1) $\CritPoint U = \Warp^{-1} (\CritPoint P)$.} 
    
    According to~\cite[Lemma 4]{Tong2023a}, it is sufficient to show that the map $Q \mapsto \Warp(Q,q)$ is everywhere a local diffeomorphism for each $q \in \indexSet$. The partial derivative of $\Warp$ with respect to its first argument is given by 
    \begin{align*}
        \frac{\partial \Warp (Q,q)}{\partial Q}  &= 
        e^{S_q \theta (Q)} \left( I + S_q Q \frac{\partial \theta (Q)}{\partial Q} \right)  .
    \end{align*}
    Using the Sherman-Morrison-Woodbury formula~\cite[Fact 3.21.3.]{Bernstein2018}, $  \det(e^{S_q \theta(Q) })  = 1$, and $ \frac{\partial \theta (Q)}{\partial Q} = 2k \nu(\epsilon)^\top$, we obtain that
    \begin{align*}
         \det \left(\frac{\partial \Warp (Q,q)}{\partial Q} \right)  &=   1 +  2k \eta (\epsilon^\top u_q)  > 0,  
    \end{align*}
    where we make use of the inequality: $2 \Abs{\eta \epsilon} \leq 1$ for all $Q \in \Sphere^3$. Therefore, the statement (S1) follows.
    
    \emph{(S2) $U$ is positive definite relative to $\SetObject$.} 
    
    Since $P$ in~\eqref{eq:basic_PF} is positive definite relative to $\SetPoint$, the zero set of $U$, denoted $\mathcal{Z}$, can be written as
    \begin{align*}
        \mathcal{Z} = \left\{(Q,q) \in \Sphere^3 \times \indexSet : \Warp (Q,q) \in \SetPoint \right\}.
    \end{align*}
    It suffices to show that $\SetObject = \mathcal{Z}$. From~\eqref{eq:warp_quat}, we have that $\Warp (\SetObject) = \SetPoint$ and thus that $\SetObject \subset \mathcal{Z}$. We next prove the other direction by contradiction. Suppose that there exists a point $(Q,q) \in \mathcal{Z} \setminus \SetObject$. Assume that $\Warp (Q,q) = \idQ$. From~\eqref{eq:warp_quat} and~\eqref{eq:warp_angle}, we have that 
    \begin{align*}
        Q &= e^{-S_q \theta(Q)} \idQ =
        \begin{bmatrix}
            \cos (\theta (Q)) \\
            - \sin (\theta (Q)) u_{q}
        \end{bmatrix}, \\
        \theta (Q) &= k \sin^2 (\theta (Q))  .
    \end{align*}   
    The last equality implies $\theta (Q) = 0$, since $k < \frac{\lambda_1}{\lambda_3} <  1$ and the inequality $x > \sin^2 (x)$ holds for all $x > 0$. Hence, $Q \in \SetPoint$, which contradicts $(Q,q) \notin  \SetObject$. Therefore, $\Warp (Q,q) \neq \idQ$. Similarly, we can show that $\Warp (Q,q) \neq -\idQ$. It follows that $\Warp (Q,q) \notin \SetPoint$, which contradicts $(Q,q) \in \mathcal{Z} $, as to prove.

    \emph{(S3) $U$ is centrally synergistic relative to $\SetPoint$ with gap exceeding some positive scalar $\delta $.} 
    
    From the very definition of the synergy gap, we have
    \begin{align*}
        \mu_U (Q,q) = \max\{0, U(Q,q) - U(Q,-q)\}.
    \end{align*}
    Therefore, $\mu_U (Q,q) > 0$ if and only if
    \begin{align}
        \mu_U (Q,q) & \stackrel{\eqref{eq:SPF_defn}}{=}  2 \left(\Gamma(Q,q) + \Gamma(Q,-q) \right) (u_q^\top A \epsilon ) \notag \\ 
        & \quad +   \left(\Gamma(Q,q)^2 - \Gamma(Q,-q)^2\right)  (u_q^\top A u_q   ) \notag \\
        & \stackrel{\eqref{eq:def_Gamma}}{=}  4 \sin(\theta(Q)) \eta \Bigl( \left(u_q^\top A \epsilon \right)   \notag \\
        & \quad + \left(\cos(\theta (Q)) - 1\right)  (u_q^\top \epsilon) \left(u_q^\top A u_q \right)
        \Bigr) . \label{eq:syngy_gap}
    \end{align}
    According to item P3) in Definition~\ref{defn:syn_defn}, $U$ has the synergism property if and only if there exists $\delta > 0$ such that $\mu_U (Q,q) > \delta$ for each $(Q,q) \in \CritPoint U \setminus \SetObject \eqqcolon \mathcal{U}$. By statement (S1), the set $\mathcal{U}$ can be written as 
    \begin{align*}
        \mathcal{U} = \bigl\{ (Q,q) & \in  \Sphere^3 \times \indexSet: \\
        & \Warp (Q,q) \in \mathcal{Q} (\pi,  v_i ),\; i=1,2,3  \bigr\} .
    \end{align*}
    
    We next evaluate the function $\mu_U$ on the set $\mc{U}$. Let $(Q,q) \in \mathcal{U}$ such that $\Warp (Q,q) \in \mathcal{Q} (\pi,   v_i ) $ for some $i \in \{1,2,3\}$. We write $\theta = \theta(Q)$ for notational convenience. It follows from~\eqref{eq:warp_quat} that 
    \begin{align}
        Q &= \pm e^{-S_q \theta } \nu (   v_i) \notag \\
        & = \pm  
        \begin{bmatrix}
            \sin (\theta ) \left(u_q^\top v_i\right) \\
            v_i + \left(\cos (\theta ) - 1\right) \left(u_q^\top v_i\right) u_q
        \end{bmatrix} = 
        \begin{bmatrix}
            \eta \\ 
            \epsilon
        \end{bmatrix}
        . \label{eq:SPF_un_crit}
    \end{align}
    Substituting~\eqref{eq:SPF_un_crit} to~\eqref{eq:warp_angle} yields
    \begin{align}
        \theta   &= k (1 - \eta^2) = k \left( 1 - \sin^2(\theta ) \left(u^\top v_i\right)^2 \right) .  \label{eq:theta_critical_point_i}
    \end{align}
    Substituting~\eqref{eq:SPF_un_crit} to~\eqref{eq:syngy_gap} yields
    \begin{align}
        & \mu_U (Q,q) = 4 \sin^2 (\theta) \left(u_q^\top v_i\right) \Bigl( u_q^\top A v_i + \notag \\ 
        & \qquad + \left(\cos (\theta) - 1\right) \left(u_q^\top v_i\right) (u_q^\top A u_q) \notag \\ 
        & \qquad   + \left(\cos(\theta ) - 1\right) \cos (\theta)  \left(u_q^\top v_i\right)  \left(u_q^\top A u_q \right)  \Bigr) \notag \\
        & \quad = 4 \sin^2 (\theta) \left(u^\top v_i\right)^2\left( \lambda_i - \sin^2 (\theta) \left(u^\top A u \right)  \right) . \label{eq:mu_critical_point_i}
    \end{align}
    Using the inequalities $u^\top v_i \neq 0$, $u^\top A u \leq \lambda_3$, $\sin^2 (\theta) \leq \sin^2 (k) \leq k < \frac{\lambda_1}{\lambda_3}$, we can conclude that for each $(Q,q) \in \Warp^{-1} (\mc{Q}(\pi, v_i))$, $\mu_U (Q,q) $ equals to a positive constant, denoted $\bar{\delta}_i$.

    Finally, if one chooses a positive scalar $\delta $ such that $\delta < \bar{\delta}_i$ for each $i \in \{1,2,3\}$, then statement (S3) follows. 
    
    \textbf{Part 3.} We next verify item 3). Substituting $u = \frac{1}{\sqrt{3}} (v_1 + v_2 + v_3)$ into~\eqref{eq:theta_critical_point_i},~\eqref{eq:mu_critical_point_i} yields
    \begin{align}
        \theta & = k \left(1 - \frac{1}{3} \sin^2(\theta)\right), \label{eq:theta_1} \\ 
        \mu(Q,q) & = \frac{4}{3} \sin^2 (\theta) \left( \lambda_i - \frac{\lambda_1 + \lambda_2 + \lambda_3}{3} \sin^2 (\theta)   \right), \label{eq:mu_1}
    \end{align}
    where $Q \in \Warp^{-1} (\mc{Q}(\pi, v_i))$. From~\eqref{eq:theta_1}, it can be seen that the warping angle $\theta$ is identical on the undesired critical points of $U$. Furthermore, we have that $ \theta \geq k (1 - \frac{1}{3} \sin^2(k)) \geq k (1 - \frac{k^2}{3}) $ and that $ \theta  \leq k $. Substituting these bounds into~\eqref{eq:mu_1}, we immediately obtain item 3), which completes the proof.
\end{pf}
\begin{rem}
    To ensure the consistency of $U$ in~\eqref{eq:SPF_defn}, it suffices to use a \emph{consistent} warping angle $\theta$. Then, the resulting hybrid feedback is also consistent.
\end{rem}
\begin{rem}
    In part 2 of the proof of Theorem~\ref{thm:cspf}, statements (S1), (S2) are mainly used to determine the undesired critical points of $U$, so as to compute the synergy gap at these points. 
\end{rem} 

\begin{rem}
    The parameter $\delta$ provided in item 3) of Theorem~\ref{thm:cspf} may be conservative. To obtain the largest upper bound of $\delta$, one can first compute the numerical solution $\theta$ of equation~\eqref{eq:theta_1}, and then evaluate the function~\eqref{eq:mu_1} at the solution with $\lambda_i = \lambda_1$. The resulting value is the desired bound.
\end{rem}

\begin{cor}\label{cor:assum}
    Consider the SPFs candidate $U$ given by~\eqref{eq:SPF_defn}. Suppose $\CritPoint U = \Warp^{-1} (\CritPoint P)$, and suppose that $U$ is positive definite relative to $\SetPoint \times \indexSet$. If $A$ has a repeated eigenvalue, then $U$ cannot be synergistic relative to $\SetPoint$.
\end{cor}
\begin{pf}
    Let $\lambda$ be the repeated eigenvalue of $A$ and $v$ be some unit eigenvector of $A$ corresponding to $\lambda$. It follows that $(Q,q) \in \Warp^{-1} (\mc{Q}(\pi, v)) $ is an undesired critical point of $U$. Let $Q =  e^{-S_q \theta } \nu (   v )$ and $\theta = \theta (Q)$. Substituting this into~\eqref{eq:syngy_gap} yields $\mu_U(Q,q) = 4 \sin^2 (\theta)  (u^\top v )^2 ( \lambda - \sin^2 (\theta)  (u^\top A u  )   )$. Since $u$ is a constant vector and the dimension of the eigenspace of $A$ corresponding to $\lambda$ is larger than one, we can assert that there exists an eigenvector $v$ such that $u^\top v = 0$ and $\mu_U$ vanishes at the critical point $(Q,q) \in \Warp^{-1} (\mc{Q}(\pi, v)) $. Therefore, synergism does not hold.
\end{pf}

\subsection{Hybrid Feedback}

Considering $U$ given by~\eqref{eq:SPF_defn},~\eqref{eq:warp_angle}, we compute the gradient $\gradA_1 U$. Define the function $\Theta: \Sphere^3 \times \indexSet \to \R^3$ as 
\begin{align}
    \Theta (Q,q) \coloneqq 
    \begin{bmatrix}
        \sin(\theta (Q)) \\ (\cos(\theta (Q)) - 1 ) u_q 
    \end{bmatrix}. \label{eq:grad_Theta}
\end{align}
From~\eqref{eq:def_Xi},~\eqref{eq:def_Gamma}, it follows that 
\begin{align}
    \gradA_1 \Gamma(Q,q) 
    &= 2k \Xi (Q,q) \nu(\epsilon) + \Theta(Q,q). \label{eq:grad_Gamma}
\end{align} 
Therefore, 
\begin{align}
    & \gradA_1  U(Q,q)  = 2 \nu(A \epsilon) + 2 \Gamma(Q,q) \nu (A u_q) \notag \\
    & \quad \qquad + 2 \left( u_q^\top A \left( \epsilon + \Gamma(Q,q) u_q \right)  \right)  \gradA_1 \Gamma(Q,q) . \label{eq:grad_U}
\end{align}
The gradient of $U$ in~\eqref{eq:grad_U} can be viewed as the gradient of $P$ in~\eqref{eq:basic_PF} augmented with a perturbation term which is linked to the logic variable. According to Theorem~\ref{thm:cspf}, the perturbation term ensures that for each $Q \in \Sphere^3 \setminus \SetPoint$, there exist $q \in \indexSet$ such that $\grad_1  U (Q,q)$ does not vanish.

The proposed hybrid feedback consists of the SPFs $U$ given by~\eqref{eq:SPF_defn},~\eqref{eq:warp_angle}, the state-feedback laws $\kappa_U$ given by~\eqref{eq:hyrid_feedback},~\eqref{eq:grad_U}, and the switching mechanism~\eqref{eq:ctr_dyn}. In comparison to the design in~\cite{Mayhew2011}, it offers the following advantages: (i) It exhibits both central synergism and consistency properties, which are highly desirable in practical applications but lacking in~\cite{Mayhew2011}. Additionally, although the state-feedback laws involve a more complex term~\eqref{eq:grad_U}, it is crucial to emphasize that the proposed feedback eliminates the requirement for the conversion mechanism of quaternion measurement (as shown in Fig.~\ref{fig:flowchart}), thanks to the consistency. Consequently, the control architecture becomes more streamlined and concise. (ii) It demonstrates a moderate switch of the state-feedback laws over jumps, as they contains the common term $\Lambda(Q)^\top \grad_1 P(Q)$, which drives the system towards the desired attitude. In contrast, the switch of the state-feedback laws in~\cite{Mayhew2011} yields completely opposite value, see Example~\ref{exmp:non_cen_SPF}. Therefore, our method can reduce chattering when significant disturbances induce unexpected jumps.

\subsection{Control Design for Problem~\ref{prob:GAS}}
Using the switching mechanism~\eqref{eq:ctr_dyn}, we propose the synergistic hybrid controller for system~\eqref{eq:plant_sys}: 
\begin{subequations} \label{eq:ctr_sys_K}
    \begin{align}
        & \tau_c = -k_p \kappa_U (Q,q) - k_d \omega, \label{eq:ctr_output_K} \\
        & 
        \begin{cases}
            \dot{q} = 0 & (Q,q) \in \mathcal{F}_K , \\
            q^+ \in \argmin_{p \in \indexSet} U(Q,p) & (Q,q) \in \mathcal{J}_K,
        \end{cases} \label{eq:ctr_dyn_K}
    \end{align}
\end{subequations}
where $k_p,k_d $ are positive gains, $\kappa_U$ is given by~\eqref{eq:hyrid_feedback} and~\eqref{eq:grad_U}, and the sets $\mc{F}_K$, $\mc{J}_K$ are defined as in~\eqref{eq:ctr_dyn}. 

Indeed, controller~\eqref{eq:ctr_dyn_K} is effective for a general SPF, regardless of whether the synergism is central or not. Therefore, the following closed-loop stability analysis proceeds with $U$ satisfying Definition~\ref{defn:syn_defn}.

Let $\mathcal{X}_2 \coloneqq \Sphere^3 \times \R^3  \times \indexSet $ and $x_2 \coloneqq (Q, \omega, q)$ denote the state space and state, respectively. Applying controller~\eqref{eq:ctr_sys_K} to system~\eqref{eq:plant_sys} yields the closed-loop system:
\begin{align}
    \mathcal{H}_2 :& 
    \begin{cases}
        \dot{x}_2 = F_2 (x_2) & x_2 \in \mathcal{F}_2 ,\\
        x_2^+ \in G_2 (x_2) & x_2 \in \mathcal{J}_2,
    \end{cases} \label{eq:clp_sys_2}
\end{align}
where the flow set and jump set are defined as $ \mathcal{F}_2 \coloneqq \{x_2 \in \mc{X}_2: \mu(Q,q) \leq \delta \}$ and $\mc{J}_2 \coloneqq \{x_2 \in \mc{X}_2: \mu(Q,q) \geq \delta \}$, respectively, and the flow map and jump map are defined as 
\begin{align*}
    F_2(x_2) &:
    \begin{cases}
        \dot{Q} = -\frac{1}{2} \Lambda(Q) \omega, \\
        J \dot{\omega} = - \omega^\times J \omega -  k_p \Lambda(Q)^\top  \grad_1 U(Q,q) - k_d \omega , \\
        \dot{q} = 0,
    \end{cases}
    \\
    G_2(x_2) & \coloneqq \left(Q, \omega, \argmin_{p \in \indexSet} U(Q,p) \right) .
\end{align*}
Define the compact sets $\mc{P}_2 , \mc{O}_2 \subset \mc{X}_2 $ as 
\begin{align}
    \mc{P}_2 &\coloneqq \{x_2 \in \mc{X}_2: (Q,q) \in \SetObject, \omega = 0\}, \\ 
    \mc{O}_2 &\coloneqq \{x_2 \in \mc{X}_2: Q \in \SetPoint, \omega = 0\},
\end{align}
where $\SetObject$ is defined as in Definition~\ref{defn:syn_defn}. The sets $\mc{P}_2 , \mc{O}_2$, and $\mc{O}_p$ satisfy the relations: $\mc{P}_2 \subset \mc{O}_2$, $\Proj_{\Sphere^3 \times \indexSet } \mc{P}_2 = \Proj_{\Sphere^3 \times \indexSet } \mc{O}_2 = \mc{O}_p$, $\Proj_{\Sphere^3 \times \indexSet }^{-1} \mc{O}_p = \mc{O}_2$. 

Now we are ready to certify the solvability of Problem~\ref{prob:GAS} by controller~\eqref{eq:ctr_sys_K}.

\begin{thm}\label{thm:stability_K}
    Problem~\ref{prob:GAS} is solvable by controller~\eqref{eq:ctr_sys_K} in the sense that the set $  \mc{O}_2 $ is robustly UGAS for system $\mc{H}_2$ given by~\eqref{eq:clp_sys_2}.
\end{thm}
\begin{extendVersion}
\begin{pf}
    The proof entails verifying statements (S1) through (S6) as follows.

    \emph{(S1) The autonomous system $\mc{H}_2$ satisfies the hybrid basic conditions.} 
    
    It suffices to show that \eqref{eq:clp_sys_2} satisfies items (A1)-(A3) of Definition~\ref{defn:basic_Hy_condition}. First, $\mc{F}_2$ and $\mc{J}_2$ are closed thanks to the continuity of $\mu_U$, which verified item (A1). Second, $F_2$ is single-valued and continuous, thereby satisfying item (A2). Finally, $G_2 $ only change the logic variable $q$ which is defined on a compact set, and so $G_2 (K) $ is compact for each compact set $ K \subset \mc{X}_2$, i.e., $G_2$ is locally bounded. Moreover, for any sequence $z_i \in \mc{X}_2$ such that $z_i \to z $ as $ i \to \infty $, the outer limit of of $G_2(z_i)$ equals the closed set $G_2 (z)$, thus $G_2$ is outer semicontinuous. This shows item (A3). 

    \emph{(S2) Each maximal solution to $\mc{H}_2$ is precompact and eventually continuous.} 
    
    Consider a Lyapunov function candidate $V_2 : \mc{X}_2 \to \R_{\geq 0}$ defined as 
    \begin{align}
        V_2(x_2) \coloneqq  U(Q,q) + \frac{1}{4 k_p} \omega^\top J \omega. \label{eq:Lyapunov}
    \end{align}
    By item P1) of Definition~\ref{defn:syn_defn}, $V_2$ is positive definite relative to $\mc{P}_2$. The change of $V_2$ at each $x_2 \in \mc{F}_2$ along flows of~\eqref{eq:clp_sys_2} is given by 
    \begin{align}
        \dot{V}_2(x_2) & = \frac{1}{2} \left(\grad_1 U(Q,q) \right)^\top \Lambda(Q)  \omega - \frac{1}{2k_p} \omega^\top \Bigl(  \omega^\times J \omega \notag \\ 
        & \quad  +  k_p \Lambda(Q)^\top \grad_1 U(Q,q) + k_d \omega \Bigr) \notag \\ 
        & = - \frac{k_d}{2k_p}\Abs{\omega}^2 \leq 0, \label{eq:flow_V_2}
    \end{align}
    where we make use of the fact that $\omega \times \omega = 0$ for all $\omega \in \R^3$. For each $x_2\in \mc{J}_2$, the change of $V_2$ over jumps of~\eqref{eq:clp_sys_2} is given by 
    \begin{align}
        \Delta V_2(x_2) &= \max_{z \in G_2 (x_2)}{V_2 (z)} - V_2(x_2) \notag \\ 
        & = -  \mu_U(Q,q) \leq -\delta_h < 0. \label{eq:jump_V_2}
    \end{align}
    Therefore, $V_2$ is nonincreasing along solutions to $\mc{H}_2$. Since $V_2$ is proper, it follows that solutions to $\mc{H}_2$ are bounded and have finitely many jumps. In addition, we make the following observations: For each $ x_2 \in \mc{F}_2 \setminus \mc{J}_2 = \{x_2 \in \mc{X}_2: \mu_{U} (Q,q) < \delta_h\}$, $F_2(x_2)$ intersects with the tangent cone to $\mc{F}_2$ at $x_2$, thanks to the openness of $\mc{F}_2 \setminus \mc{J}_2$; $G_2(\mc{J}_2) \subset \mc{F}_2 \setminus \mc{J}_2 \subset \mc{X}_2$. Combining these observations, we have established the statement (S2) by invoking Proposition~\ref{prop:existence_condition}.

    \emph{(S3) The set $\mc{P}_2$ is UGAS for $\mc{H}_2$.} 
    
    First, from~\eqref{eq:flow_V_2},~\eqref{eq:jump_V_2}, we conclude that the set $\mc{P}_2$ is stable as per~\cite[Theorem 3.19]{Sanfelice2021}. Recall the notation $\dot{V}_2^{-1} (0) \coloneqq \{x_2 \in \mc{F}_2 : \dot{V}_2(x_2) = 0\}$. Applying the hybrid invariance principle~\cite[Theorem 3.23]{Sanfelice2021}, we have that precompact solutions to $\mc{H}_2$ converge to the largest weakly invariant set $\mc{I}_2 \subset V_2^{-1} (r) \bigcap\dot{V}_2^{-1} (0) $ for some $r \in \R$. 
    
    For each $x_2 \in \dot{V}_2^{-1} (0) $, we make the following observations: From~\eqref{eq:flow_V_2},~\eqref{eq:clp_sys_2},~\eqref{eq:hyrid_feedback}, and Lemma~\ref{lem:kappa_crit}, it follows that $\omega = 0$, $\grad_1 U(Q,q) = 0$, and $(Q,q) \in \CritPoint U$. On the other hand, $x_2 \in \mc{F}_2$ implies $\mu(Q,q) \leq \delta_h \leq \delta$, and consequently, we obtain from item P3) of Definition~\ref{defn:syn_defn} that $(Q,q) \notin \CritPoint U \setminus \SetObject$. From item P2) of Definition~\ref{defn:syn_defn}, it follows that $\SetObject \subset \CritPoint U$ and that $(Q,q) \in \SetObject  $.
    
    We therefore obtain $ \dot{V}_2^{-1} (0) = \mc{P}_2$, and hence $\mc{I}_2 \subset V_2^{-1} (0) \bigcap \dot{V}_2^{-1} (0) = \mc{P}_2$. It follows that the set $\mc{P}_2$ is globally attractive and thus globally asymptotically stable for $\mc{H}_2$. Thanks to statement (S1), the set $\mc{P}_2$ is UGAS for $\mc{H}_2$ as per~\cite[Theorem 7.12]{Goebel2012}.

    \emph{(S4) The set $\mc{O}_2$ is UGAS for $\mc{H}_2$.}
    
    According to~\cite[Propositin 7.5 \& Theorem 7.12]{Goebel2012}, it suffices to verify that $\mc{O}_2$ is forward invariant, due to $\mc{P}_2 \subset \mc{O}_2$. For each maximal solution $\phi$ to $\mc{H}_2$, we make the following observation: If $\phi(0,0) \in \mc{P}_2 \subset \mc{F}_2 \setminus \mc{J}_2$, $\phi$ remains unchanged in $\mc{P}_2$, since $F_2$ vanishes on $\mc{P}_2$; If $\phi(0,0) \in \mc{O}_2 \setminus \mc{P}_2 \subset \mc{J}_2 \setminus \mc{F}_2$, $\phi$ takes a single immediate jump into the set $\mc{P}_2$, which is due to $G_2 (\mc{O}_2) = \mc{P}_2$. Therefore, $\mc{O}_2$ is forward invariant, as to prove.

    \emph{(S5) The set $\mc{O}_2$ is robustly UGAS for $\mc{H}_2$.} 
    
    From statements (S1), (S4) and~\cite[Theorem 7.21]{Goebel2012}, we assert that the set $\mc{O}_2$ is UGpAS for $\mc{H}_{2,\rho}$, the $\rho$-perturbation of $\mc{H}_2$, where $\rho : \mc{X}_2 \to \R_{\geq 0}$ is a continuous function that is positive on $\mc{X}_2 \setminus \mc{O}_2 $. According to~\cite[Propositin 6.28]{Goebel2012}, $\mc{H}_{2,\rho}$ satisfies the hybrid basic conditions. From~\eqref{eq:perturb_HySys}, we have that $\mc{F}_2 \subset \mc{F}_{2,\rho} \subset \mc{X}_2$ and $\mc{J}_2 \subset \mc{J}_{2,\rho} \subset \mc{X}_2$, which yields $\mc{F}_{2,\rho} \bigcup \mc{J}_{2,\rho} = \mc{X}_2$ due to $\mc{X}_2 = \mc{F}_2 \bigcup \mc{J}_2 $. Combining these set relations yields 
    $\mc{F}_{2,\rho} \setminus \mc{J}_{2,\rho} = \mc{F}_{2,\rho} \bigcap (\mc{X}_2 \setminus \mc{J}_{2,\rho}) =  \mc{F}_{2,\rho}  \bigcap ( (\mc{F}_2 \bigcup \mc{J}_2) \setminus \mc{J}_{2,\rho})  = \mc{F}_{2,\rho}  \bigcap ( (\mc{F}_2 \setminus \mc{J}_2 ) \setminus \mc{J}_{2,\rho})  = (\mc{F}_2 \setminus \mc{J}_2) \setminus \mc{J}_{2,\rho} $. Since the set $\mc{F}_2 \setminus \mc{J}_2$ is open and the set $\mc{J}_{2,\rho}$ is closed, it follows that the set $\mc{F}_{2,\rho} \setminus \mc{J}_{2,\rho}$ is open. Then, we make the following observations: For each $ x_2 \in \mc{F}_{2,\rho} \setminus \mc{J}_{2,\rho} $, $F_{2,\rho}(x_2)$ intersects with the tangent cone to $\mc{F}_{2,\rho}$ at $x_2$; Solutions to $\mc{H}_{2,\rho}$ cannot blow up to infinity at a finite time, due to the UGpAS property of the compact set $\mc{O}_2$; $G_{2,\rho}(\mc{J}_{2,\rho}) \subset \mc{F}_{2,\rho} \bigcup \mc{J}_{2,\rho}$. According to these observations, we can conclude that maximal solutions to $\mc{H}_{2,\rho}$ are complete as per Proposition~\ref{prop:existence_condition}, and statement (S5) follows. This completes the proof. 
\end{pf} 
\end{extendVersion}

\begin{rem}
    Theorems~\ref{thm:stability},~\ref{thm:stability_K} present enhanced robustness results compared to the previous work in~\cite{Mayhew2011,Tong2023a}. The key difference lies in the fact that the desired sets $\mathcal{O}_1$ and $\mc{O}_2$ are proven to be robustly UGAS rather than robustly UGpAS. This means that there exits the perturbed closed system whose solutions can evolve for arbitrarily long hybrid time and approach to the desired set, thereby making the introduced perturbation more practical and reasonable in real-world scenarios.
\end{rem}

\section{Simulations} \label{sec:sim}
This section presents the simulations of Problem~\ref{prob:GAS} with controller~\eqref{eq:ctr_sys_K} that are generated via the hybrid equations toolbox for MATLAB~\cite{Sanfelice2013}. The proposed centrally synergistic hybrid controller is referred to as ``CSH'', which is generated from the SPFs in~\eqref{eq:SPF_defn}. For comparison, the noncentrally synergistic hybrid controller in~\cite{Mayhew2011} (see Example~\ref{exmp:non_cen_SPF}) is considered and referred to as ``NCSH''. We also consider the continuous feedback, i.e., these synergistic hybrid controllers with fixed logic variable. The CSH controller with fixed logic variable $q = 1$ is referred to as ``CS $q=1$''. No conversion mechanism is used to select quaternion from attitude measurements.

The inertial matrix of the rigid-body system~\eqref{eq:dyn_quat}-\eqref{eq:dyn_angVel} is assumed to be $J = \operatorname{diag}([6.4,6.7,9.3])$. The parameters of the CSH controller are set as follows: $A = \operatorname{diag}([0.6,0.8,1])$, $u = \frac{1}{\sqrt{3}}[1,1,1]^\top$, $k = 0.54$, $\delta_h = 0.1$. The parameters of the NCSH controller are set to be $\delta_h = 0.1$. The positive gains of controller~\eqref{eq:ctr_sys_K} are chosen to be $k_p = 30$ and $k_d = 15$. 

\begin{figure}[htb]
    \centering
    \includegraphics[width=8cm]{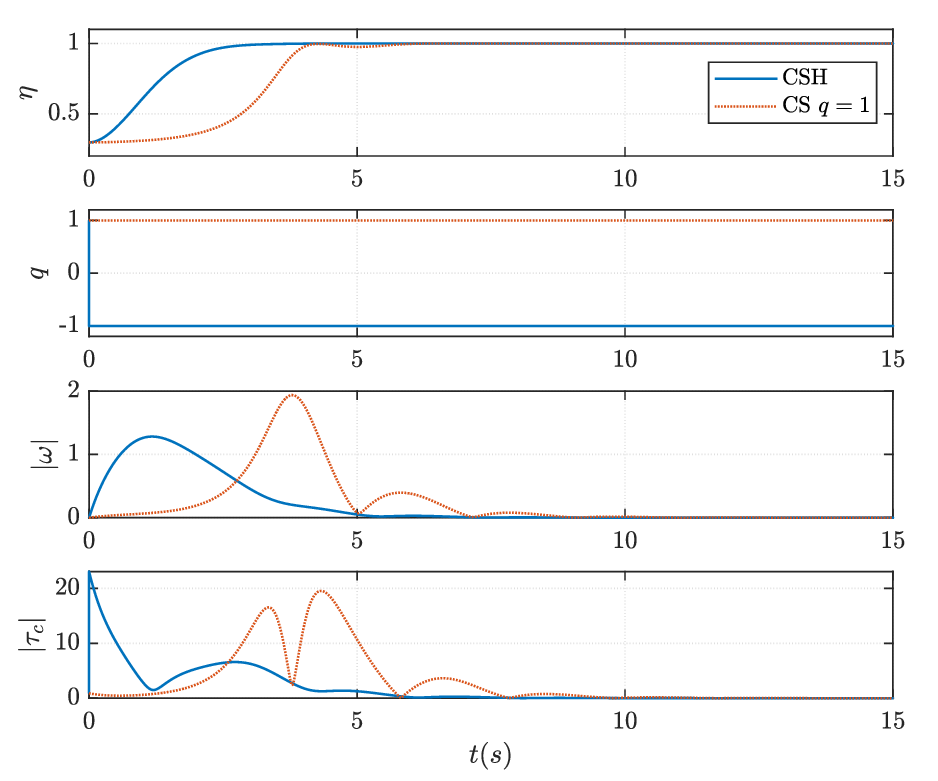}
    \caption{Simulation 1: CSH versus CSH $q=1$. The initial conditions: $Q(0) = [0.297, -0.028, 0.013, 0.954]^\top$, $\omega(0) = [0,0,0]^\top$, $q(0) = 1$.}
    \label{fig:SecondOrderGlobal_231204}
\end{figure}

\begin{figure}[htb]
    \centering
    \includegraphics[width=8cm]{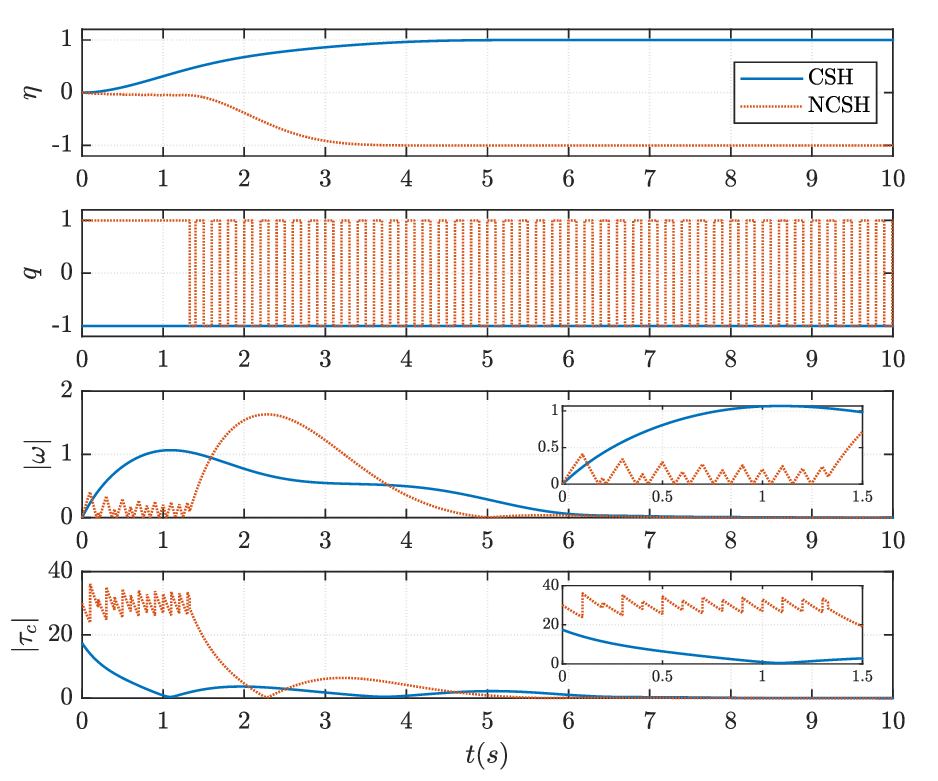}
    \caption{Simulation 2: CSH versus NCSH. The initial conditions: $Q(0) = [0, 0.6, 0.8, 0]^\top$, $\omega(0) = [0,0,0]^\top$, $q(0) = -1$ for CSH, and $q(0) = 1$ for NCSH.}
    \label{fig:SecondOrderInconsistent_231204}
\end{figure}

\begin{figure}[htb]
    \centering
    \includegraphics[width=8cm]{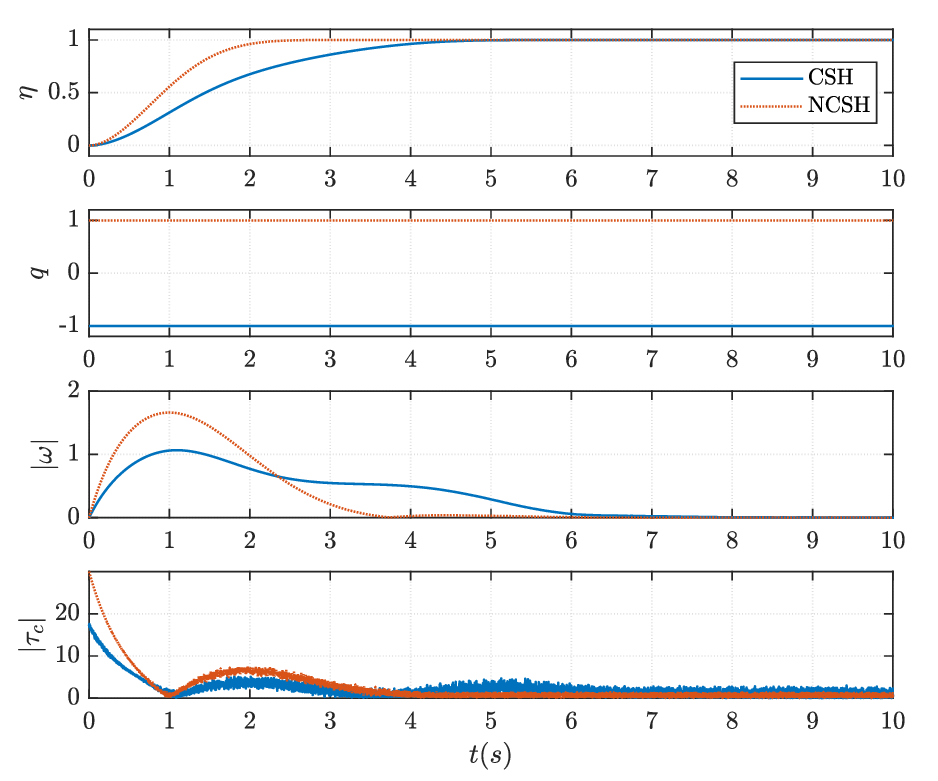}
    \caption{Simulation 3: CSH versus NCSH. The initial conditions: $Q(0) = [0, 0.6, 0.8, 0]^\top$, $\omega(0) = [0,0,0]^\top$, $q(0) = -1$ for CSH, and $q(0) = 1$ for NCSH.}
    \label{fig:SecondOrder_AddNoise_05_231204}
\end{figure}

\begin{figure}[htb]
    \centering
    \includegraphics[width=8cm]{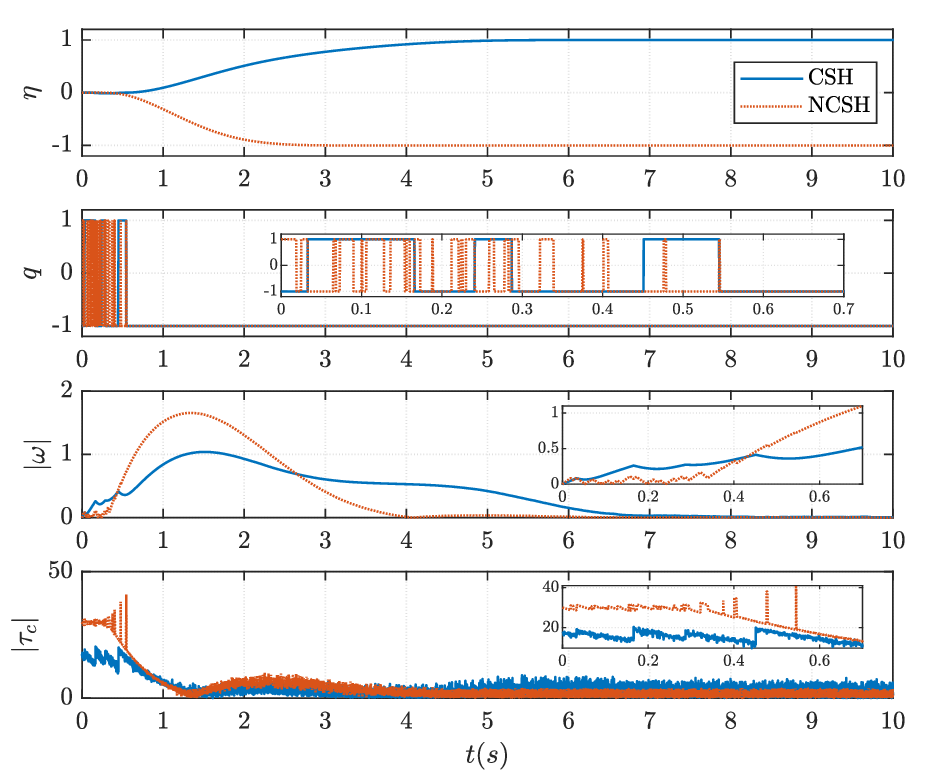}
    \caption{Simulation 4: CSH versus NCSH. The initial conditions: $Q(0) = [0, 0.6, 0.8, 0]^\top$, $\omega(0) = [0,0,0]^\top$, $q(0) = -1$ for CSH, and $q(0) = 1$ for NCSH.}
    \label{fig:SecondOrder_AddNoise_13_231204}
\end{figure}

Four numerical simulations are illustrated in Figs.~\ref{fig:SecondOrderGlobal_231204}-\ref{fig:SecondOrder_AddNoise_13_231204}, respectively. Each figure displays the time histories of the scalar part $\eta(t)$ of the rigid-body attitude quaternion $Q(t)$, the logic variable $q(t)$, the Euclidean norm of the angular velocity $\omega(t)$, and the Euclidean norm of the torque input $\tau_c(t)$.

\emph{Simulation~1: global convergence of the CSH controller.} Due to the lack of an analytic solution for the undesired critical points of the function $U$ in~\eqref{eq:SPF_defn}, we used a numerical approximation of the undesired equilibrium point with $q = 1$ as initial conditions. As shown in Fig.~\ref{fig:SecondOrderGlobal_231204}, the torque input of the CS $q = 1$ controller was initially vanishing, resulting in a significantly longer time to reach the desired attitude. In contrast, the CSH controller detected this situation and judiciously switched the nonvanishing feedback indexed by $q = -1$. It is important to note that if the initial conditions were exactly set at the undesired equilibrium point with $q = 1$, CS $q = 1$ controller would produce zero torque, preventing convergence to the desired attitude.

\emph{Simulation 2: effects of memoryless quaternion measurements.} The quaternion measurement was defined as $Q_m = sQ$, where $s(t)$ represents a square wave with a frequency of $5\mathrm{Hz}$. It is shown in Fig.~\ref{fig:SecondOrderInconsistent_231204} that the NCSH controller exhibited chattering near the initial conditions $\eta = 0$. This behavior occurred due to the frequent alterations of the controller's intended rotation direction caused by the discontinuities in quaternion measurements. However, when the attitude $Q$ was far from $\eta = 0$, the NCSH controller detected the incorrect selection of quaternion measurement by using switching mechanism and therefore adjusted the sign of its feedback by modifying the index logic. Conversely, the proposed CSH controller remained unaffected by the discontinuous nature of the quaternion measurements thanks to its consistency.

\emph{Simulations~3 and~4: robustness to measurement noises.} In line with the simulation setup in~\cite{Mayhew2011}, the quaternion measurement is defined as $Q_m = (Q + n e)/ |Q+ne|$, where $e = v/|v|$ and $v \in \R^4$ is drawn from a zero-mean Gaussian distribution with unit variance, and $n$ is drawn from a uniform distribution over the interval $[0,n_m]$. In Simulation 3, we set $n_m = 0.05$ to represent a scenario with small noise, while in Simulation 4, we set $n_m = 0.13$ to simulate a scenario with large noise. Fig.~\ref{fig:SecondOrder_AddNoise_05_231204} illustrates the robustness of both CSH and NCSH controllers to measurement noises to some extent. However, when the measured quaternions erroneously fell within the jump sets, as depicted in Fig.~\ref{fig:SecondOrder_AddNoise_13_231204}, the switching mechanism resulted in chattering of the angular velocity. Notably, the NCSH controller exhibited higher sensitivity to large noises. This can be attributed to the fact that the NCSH controller employs two opposing state-feedback laws, and each jump significantly alters the torque, leading to a more aggressive response to frequent switching. In contrast, the state-feedback laws of the CSH controller share a common action term as shown in~\eqref{eq:grad_U}, resulting in a slightly reduced chattering effect caused by improper switching.

\section{Conclusion} \label{sec:Conclusion}
We address the attitude stabilization problem for a rigid-body system using quaternion-based synergistic hybrid feedback. Our control law is based on a new family of synergistic potential functions, leading to robust uniform global asymptotic stability. It exhibits two key advantages over existing results. Firstly, the central synergism and consistency properties are achieved by employing the minimal number of potential functions. Secondly, it enhances robustness by establishing the completeness of solutions to the perturbed closed-loop system. In the future, further exploration can be done by combining our proposed approach with disturbance rejection techniques.

\appendix

\section{Proof of Example~\ref{exmp:basic_PF}}
From the very definition of the gradient, we have that $\gradA P (Q) = [0,2  \epsilon^\top A]^\top $ and that 
\begin{align*}
    \Abs{\grad P (Q)}^2 &= (\gradA P (Q))^\top \Pi (Q) \gradA P (Q) \\
    &= 4 \epsilon^\top A \left( I_3 - \epsilon \epsilon^\top  \right) A \epsilon,
\end{align*}
where we have used the fact that $\Pi (Q)^2 = \Pi (Q) $ for all $Q \in \Sphere^3 $. Therefore, 
\begin{align*} 
    \CritPoint P &= \left\{ Q = [\eta,\epsilon^\top]^\top \in \Sphere^3 :  \epsilon^\top A \left( I_3 - \epsilon \epsilon^\top  \right) A \epsilon = 0\right\}.
\end{align*}
Obviously, any $Q \in \Sphere^3 $ with $ \epsilon  = 0$ is contained in $ \CritPoint P$, i.e., $\SetPoint \subset \CritPoint P$. For $Q \in \CritPoint P$ with $\Abs{\epsilon} \neq 0$, we must have that $\Abs{\epsilon} = 1$ (otherwise, $\Abs{\epsilon} < 1$ and thus $  A ( I_3 - \epsilon \epsilon^\top ) A $ is positive definite), and hence, that $\epsilon$ equals a scalar multiple of $A \epsilon$, i.e., $\epsilon$ is a unit eigenvector of $A$.

\begin{extendVersion}
\section{Analysis Tools} \label{sec:Tools}
\begin{defn}[{\cite[Assumption 6.5]{Goebel2012}}] \label{defn:basic_Hy_condition}
    The hybrid system~\eqref{eq:HySys} is said to satisfy the \emph{hybrid basic conditions} if the following hold:
    \begin{enumerate}[({A}1)]
        \item $\mathcal{F}$ and $\mc{J}$ are closed subsets of $\R^n$;
        \item $F$ is outer semicontinuous and locally bounded relative to $\mc{F}$, $\mc{F} \subset \domain F$, and $F(x)$ is convex for every $x \in \mc{F}$;
        \item $G$ is outer semicontinuous and locally bounded relative to $\mc{J}$, and $\mc{J} \subset \domain G$.
    \end{enumerate}
\end{defn}

\begin{prop}[{\cite[Propositin 6.10]{Goebel2012}}] \label{prop:existence_condition}
    Consider system given by~\eqref{eq:HySys} and suppose it satisfy the hybrid basic conditions. Take an arbitrary $x_0 \in \mc{F} \bigcup \mc{J}$. If $x_0 \in \mc{J}$ or 
    \begin{enumerate}[{(VC)}]
        \item there exists a neighborhood $\mc{U}$ of $x_0$ such that for each $x \in \mc{U} \bigcap \mc{F}$,\footnote{$T_{\mc{F}} (x)$ denotes the tangent cone of the set $\mc{F}$ at the point $x$.} 
        \begin{align*}
            F(x) \bigcap T_{\mc{F}} (x)\neq \emptyset, 
        \end{align*}
    \end{enumerate}
    then there exists a nontrivial solution $\phi$ to $\mc{H}$ with $\phi(0,0) = x_0$. If (VC) holds for every $x_0 \in \mc{F} \setminus \mc{J}$, then there exists a nontrivial solution to $\mc{H}$ from every initial point in $ \mc{F} \bigcup \mc{J}$, and every maximal solution $\phi $ to $\mc{H}$ satisfies exactly one of the following conditions:
    \begin{enumerate}[(a)]
        \item $\phi$ is complete:
        \item $\domain \phi$ is bounded and the interval $I^J$, where $J = \sup_j \domain \phi$, has nonempty interior and $t \mapsto \phi(t,J)$ is a maximal solution to $\dot{z} \in F(z)$, $z \in \mc{F}$ satisfying $\lim_{t \to T} \Abs{\phi(t,J)} = \infty$, where $T = \sup_t \domain \phi$;
        \item $\phi(T,J) \notin \mc{F} \bigcup \mc{J}$, where $(T,J) = \sup  \domain \phi$
    \end{enumerate}
    Furthermore, if $G(\mc{J}) \subset \mc{F} \bigcup \mc{J} $, then (c) above does not occur.
\end{prop}
\end{extendVersion}

\bibliographystyle{unsrt}
\bibliography{References/references}

\begin{thebibliography}{10}

\bibitem{Basso2023}
Erlend~A. Basso, Henrik~M. Schmidt-Didlaukies, Kristin~Y. Pettersen, and Asgeir~J. S{\o}rensen.
\newblock Global asymptotic tracking for marine vehicles using adaptive hybrid feedback.
\newblock {\em {IEEE} Transactions on Automatic Control}, 68(3):1584--1599, mar 2023.

\bibitem{Shao2023}
Xiaodong Shao, Qinglei Hu, Yang Shi, and Youmin Zhang.
\newblock Fault-tolerant control for full-state error constrained attitude tracking of uncertain spacecraft.
\newblock {\em Automatica}, 151:110907, may 2023.

\bibitem{Barman2023}
Saumitra Barman and Manoranjan Sinha.
\newblock Satellite attitude control using double-gimbal variable-speed control moment gyroscope: Single-loop control formulation.
\newblock {\em Journal of Guidance, Control, and Dynamics}, pages 1--17, 2023.

\bibitem{Bhat2000}
Sanjay~P. Bhat and Dennis~S. Bernstein.
\newblock A topological obstruction to continuous global stabilization of rotational motion and the unwinding phenomenon.
\newblock {\em Systems \& Control Letters}, 39(1):63--70, 2000.

\bibitem{Maithripala2006}
D.~H.~S. Maithripala, J.~M. Berg, and W.~P. Dayawansa.
\newblock Almost-global tracking of simple mechanical systems on a general class of lie groups.
\newblock {\em {IEEE} Transactions on Automatic Control}, 51(2):216--225, 2006.

\bibitem{Akhtar2021}
Adeel Akhtar and Steven~L. Waslander.
\newblock Controller class for rigid body tracking on {SO(3)}.
\newblock {\em {IEEE} Transactions on Automatic Control}, 66(5):2234--2241, may 2021.

\bibitem{Tejaswi2023}
Tejaswi K.C. and Sukumar Srikant.
\newblock Attitude control via a feedback integrator based observer.
\newblock {\em Automatica}, 151:110882, may 2023.

\bibitem{Shuster1993}
Malcolm~D. Shuster.
\newblock A survey of attitude representations.
\newblock {\em Journal of the Astronautical Sciences}, 41(4):439--517, 1993.

\bibitem{Mayhew2013}
Christopher~G. Mayhew, Ricardo~G. Sanfelice, and Andrew~R. Teel.
\newblock On path-lifting mechanisms and unwinding in quaternion-based attitude control.
\newblock {\em {IEEE} Transactions on Automatic Control}, 58(5):1179--1191, may 2013.

\bibitem{Sanfelice2006}
R.~G. Sanfelice, M.~J. Messina, S.~Emre Tuna, and A.~R. Teel.
\newblock Robust hybrid controllers for continuous-time systems with applications to obstacle avoidance and regulation to disconnected set of points.
\newblock In {\em 2006 American Control Conference}, pages 3352--2257, 2006.

\bibitem{Mayhew2011}
Christopher~G. Mayhew, Ricardo~G. Sanfelice, and Andrew~R. Teel.
\newblock Quaternion-based hybrid control for robust global attitude tracking.
\newblock {\em {IEEE} Transactions on Automatic Control}, 56(11):2555--2566, 2011.

\bibitem{Mayhew2013a}
Christopher~G. Mayhew and Andrew~R. Teel.
\newblock Synergistic hybrid feedback for global rigid-body attitude tracking on {SO}(3).
\newblock {\em {IEEE} Transactions on Automatic Control}, 58(11):2730--2742, 2013.

\bibitem{Lee2015}
Taeyoung Lee.
\newblock Global exponential attitude tracking controls on {SO(3)}.
\newblock {\em {IEEE} Transactions on Automatic Control}, 60(10):2837--2842, 2015.

\bibitem{Mayhew2011b}
Christopher~G. Mayhew and Andrew~R. Teel.
\newblock Synergistic potential functions for hybrid control of rigid-body attitude.
\newblock In {\em Proceedings of the 2011 American Control Conference}, pages 875--880, 2011.

\bibitem{Berkane2017a}
Soulaimane Berkane and Abdelhamid Tayebi.
\newblock Construction of synergistic potential functions on {SO}(3) with application to velocity-free hybrid attitude stabilization.
\newblock {\em {IEEE} Transactions on Automatic Control}, 62(1):495--501, jan 2017.

\bibitem{Berkane2017b}
Soulaimane Berkane, Abdelkader Abdessameud, and Abdelhamid Tayebi.
\newblock Hybrid global exponential stabilization on {SO(3)}.
\newblock {\em Automatica}, 81:279--285, 2017.

\bibitem{Casau2015}
Pedro Casau, Ricardo~G. Sanfelice, Rita Cunha, and Carlos Silvestre.
\newblock A globally asymptotically stabilizing trajectory tracking controller for fully actuated rigid bodies using landmark-based information.
\newblock {\em International Journal of Robust and Nonlinear Control}, 25(18):3617--3640, 2015.

\bibitem{Tong2023}
Xin Tong and Shing~Shin Cheng.
\newblock Synergistic potential functions from single modified trace function on {SO}(3).
\newblock {\em Automatica}, 154:111070, aug 2023.

\bibitem{Gui2018}
Haichao Gui and Anton H.~J. de~Ruiter.
\newblock Global finite-time attitude consensus of leader-following spacecraft systems based on distributed observers.
\newblock {\em Automatica}, 91:225--232, 2018.

\bibitem{Huang2021}
Yi~Huang and Ziyang Meng.
\newblock Global finite-time distributed attitude synchronization and tracking control of multiple rigid bodies without velocity measurements.
\newblock {\em Automatica}, 132:109796, 2021.

\bibitem{Invernizzi2022}
Davide Invernizzi, Marco Lovera, and Luca Zaccarian.
\newblock Global robust attitude tracking with torque disturbance rejection via dynamic hybrid feedback.
\newblock {\em Automatica}, 144:110462, October 2022.

\bibitem{Schlanbusch2015}
Rune Schlanbusch and Esten Ingar~Ingar Grotli.
\newblock Hybrid certainty equivalence control of rigid bodies with quaternion measurements.
\newblock {\em {IEEE} Transactions on Automatic Control}, 60(9):2512--2517, sep 2015.

\bibitem{Zhang2022}
Dandan Zhang, Xin Jin, and Hongye Su.
\newblock Robust global attitude control: Random reset rule.
\newblock {\em {IEEE} Transactions on Automatic Control}, pages 1--8, 2022.

\bibitem{Espindola2023}
Eduardo Esp{\'{\i}}ndola and Yu~Tang.
\newblock A four-{DOF} lagrangian approach to attitude tracking.
\newblock {\em Automatica}, 151:110880, may 2023.

\bibitem{Tong2023a}
Xin Tong and Shing~Shin Cheng.
\newblock Global stabilization of antipodal points on n-sphere with application to attitude tracking.
\newblock {\em IEEE Transactions on Automatic Control}, pages 1--8, 2023.

\bibitem{Goebel2012}
Rafal Goebel, Ricardo~G. Sanfelice, and Andrew~R. Teel.
\newblock {\em Hybrid Dynamical Systems: Modeling, Stability, and Robustness}.
\newblock Princeton University Press, New Jersey, 2012.

\bibitem{Sanfelice2021}
Ricardo~G. Sanfelice.
\newblock {\em Hybrid Feedback Control}.
\newblock Princeton University Press, New Jersey, 2021.

\bibitem{Gallier2003}
Jean Gallier and Dianna Xu.
\newblock Computing exponentials of skew-symmetric matrices and logarithms of orthogonal matrices.
\newblock {\em International Journal of Robotics and Automation}, 18(1):10--20, 2003.

\bibitem{Bernstein2018}
Dennis~S. Bernstein.
\newblock {\em Scalar, Vector, and Matrix Mathematics: Theory, Facts, and Formulas}.
\newblock Princeton University Press, Princeton, New Jersey, revised and expanded edition, 2018.

\bibitem{Sanfelice2013}
Ricardo~G. Sanfelice, David~A. Copp, and Pablo Nanez.
\newblock A toolbox for simulation of hybrid systems in matlab/simulink: Hybrid equations ({HyEQ}) toolbox.
\newblock In {\em Proceedings of the 16th International Conference on Hybrid Systems: Computation and Control}, HSCC '13, page 101–106, New York, NY, USA, 2013. Association for Computing Machinery.

\end{thebibliography}

\end{document}